\documentclass[conference]{IEEEtran}
\IEEEoverridecommandlockouts
\usepackage{cite}
\usepackage{amsmath,amssymb,amsfonts}
\usepackage{graphicx}
\usepackage{textcomp}
\usepackage{xcolor}
\usepackage{mathrsfs} % 提供 \mathscr 符号
\usepackage{algorithm}
\usepackage{algpseudocode} % For additional formatting if needed
\def\BibTeX{{\rm B\kern-.05em{\sc i\kern-.025em b}\kern-.08em
    T\kern-.1667em\lower.7ex\hbox{E}\kern-.125emX}}
\begin{document}

\title{Joint Routing and Control Optimization in VANET
\thanks{ }
}

\author{\IEEEauthorblockN{1\textsuperscript{st} Chen Huang}
\IEEEauthorblockA{\textit{School of Cyber Engineering} \\
\textit{Xidian University}\\
Xi'an, China }
\and
\IEEEauthorblockN{2\textsuperscript{nd} Dingxuan Wang}
\IEEEauthorblockA{\textit{School of Cyber Engineering} \\
\textit{Xidian University}\\
Xi'an, China}
\and
\IEEEauthorblockN{3\textsuperscript{rd} Ronghui Hou}
\IEEEauthorblockA{\textit{School of Cyber Engineering} \\
\textit{Xidian University}\\
Xi'an, China}
% \and
% \and
% \and
}

\maketitle

\begin{abstract}
% In highly dynamic vehicular networks, ensuring safe vehicle driving and improving the performance of intelligent transportation systems, such as throughput, is crucial. This paper designs a joint optimization problem that comprehensively addresses two key requirements: platoon control and data transmission. By leveraging effective platoon safety control, continuous vehicle movement is guaranteed, while trajectory optimization guides vehicles in selecting optimal nodes for data transmission. This ensures the real-time and reliable transmission of data across various traffic scenarios and further enhances the stability of platoon control through more accurate data transmission, thereby improving the overall performance of the intelligent transportation system. Simulations conducted in multiple complex vehicular network scenarios—such as straight roads and lane merging—demonstrate that the proposed scheme achieves excellent control and transmission performance.
In this paper, we introduce DynaRoute, an adaptive joint optimization framework for dynamic vehicular networks that simultaneously addresses platoon control and data transmission through trajectory-aware routing and safety-constrained vehicle coordination. DynaRoute guarantees continuous vehicle movement via platoon safety control with optimizing transmission paths through real-time trajectory prediction and ensuring reliable data. Our solution achieves three key objectives: (1) maintaining platoon stability through accurate data transmission, (2) enabling adaptive routing based on vehicle movement patterns, and (3) enhancing overall intelligent transportation system performance. DynaRoute equires predefined traffic models and adapts to dynamic network conditions using local vehicle state information. We present comprehensive simulation results demonstrating that DynaRoute maintains control and transmission performance in multiple complex scenarios while significantly improving throughput and reliability compared to traditional approaches.
\end{abstract}


\section{Introduction}
\par The rapid advancement of Internet of Things (IoT) technology has transformed modern vehicles from simple transportation tools into sophisticated communication hubs. Functioning as both receiving and transmitting nodes, these connected vehicles play a dual role in Intelligent Transportation Systems (ITS). As receivers, they continuously collect real-time environmental data, enabling autonomous trajectory planning and adaptive velocity control through intelligent algorithms, thereby enhancing road throughput and optimizing overall ITS performance. Simultaneously, as transmitters, they leverage roadside infrastructure to deliver essential services including navigation, communication, and entertainment \cite{ref1}. 
\par The high mobility of vehicles leads to rapid topology changes and unstable connections, particularly in congested areas where bandwidth exhaustion leads to packet backlog, preventing real-time transmission and introducing additional latency. The shared wireless channel experiences frequent collisions, further exacerbating performance degradation caused by data transmission delays, with Dedicated Short-Range Communications (DSRC) being especially vulnerable to interference and signal degradation \cite{ref2,ref3}. These dynamics significantly impair IoV communications, safety information delivery and platoon coordination. Traditional routing methods often were proved inadequate in addressing these mobility-induced challenges \cite{ref4}, highlighting the need for solutions that better accommodate dynamic network conditions while optimizing routing overhead to improve resource utilization in increasingly complex IoT environments.
% \par Control theory is a branch of applied mathematics that studies how to manipulate system behaviors to achieve desired outputs through feedback loops and dynamic adjustments. It is widely used in autonomous vehicles, robotics, industrial automation, and communication systems. Recent research has developed various security control strategies (optimal control, MPC, nonlinear methods) to address mobility challenges in IoV, enabling optimal vehicle trajectories and stable link identification through movement prediction. Fortunately, recent advances in joint optimization of vehicular communication and control systems offer promising pathways to address these challenges\cite{ref5,ref6,ref7}. However, existing joint optimization methods primarily focus on transmission power allocation in wireless channels, while we focus on intelligent path selection which is equally vital for performance optimization. Effective path management in such networks requires a comprehensive evaluation of multiple factors and the implementation of adaptive, intelligent strategies.
\par With the growing demand for intelligent transportation systems, autonomous vehicle operation relies critically on control mechanisms to ensure safe driving. Many works simultaneously consider the communication system and the control system in intelligent vehicular systems. Some works study the impact of communication on control performance to find the appropriate communication parameters \cite{ref11,ref12,ref13,ref14,ref15}. Other works design the communication schemes to meet the control system requirements\cite{ref5,ref6,ref7,ref8,ref9,ref10}. In this paper, we study the joint optimization of routing and control. In one hand, efficient routing scheme produces smaller packet transmission delay, and then, the control system may obtain more real-time state information to perform a better control. On the other hand, the control system timely feedbacks the controlling decision to help the routing system adapts its routing solution to improve network communication quality. Although a very related work in \cite{ref16} also jointly considered routing and controlling, the work did not simultaneously optimization both systems but optimized both systems individually. In this paper, we propose DynaRoute—an integrated control-communication optimization framework that constructs mobility-aware link evaluation metrics and accurately predicts vehicle trajectories. Our solution dynamically optimizes transmission routes through real-time node position prediction, ensuring reliable and timely data delivery across diverse traffic conditions while simultaneously improving platoon control performance through enhanced transmission precision. Compared to conventional approaches, the proposed method demonstrates significant improvements in network efficiency and robustness under high-mobility scenarios, while advancing the overall performance of intelligent transportation systems.
\section{Related work}
\par In Vehicular Ad-hoc Network (VANET), inter-vehicle communication is affected by various environmental and nodal factors, which further impact vehicle platoon and overall system performance. Consequently, maintaining ideal inter-vehicle spacing, reducing spacing errors, and increasing communication reliability have become new research objectives. The performance of vehicle platoon control depends on the topology and quality of wireless communications. \cite{ref5} investigated a constant time headway-based platoon control mechanism under limited communication range and random packet loss conditions. \cite{ref6} studied the longitudinal control problem of vehicles based on a uniform constant time headway strategy, theoretically analyzing the upper bound of time delay to ensure platoon stability under packet loss conditions. To better address data transmission under constrained communication resources, \cite{ref7} proposed a dynamic event-triggered mechanism to mitigate the impact of network topology on control performance in VANET. \cite{ref8} designed a dynamic event-triggered mechanism for vehicles that considers unknown external disturbances and uncertain control inputs. This mechanism adaptively adjusts communication based on vehicle state variations and bandwidth availability, thereby improving communication performance. A platoon control framework was designed to maintain robust control performance even under random network topology variations. \cite{ref9} examined the transmission mechanism of Cooperative Awareness Messages (CAMs) and developed a sampled-data feedback controller to eliminate the effects of random packet loss and external disturbances. For Vehicular Cyber-Physical Systems (VCPS), \cite{ref10} considered the beamforming scheme on the control performance, and formated the control optimization with the constraints of transmit power and beamwidth. By leveraging network topology information as a parameter, the control performance was further optimized. These studies comprehensively consider the impact of data communication on vehicle platoon control, demonstrating various approaches to enhance control performance under challenging communication conditions. 
\par Current research has addressed these challenges through co-design approaches that jointly consider communication and control problems. \cite{ref11} derived the transmission delay allowed to guarantee a required control performance. \cite{ref12} and \cite{ref13} found appropriate transmission scheduling schemes  using the node mobility information which is provided by the node control function, while the works did not consider the impact of data transmission on control performance. \cite{ref14} firstly found a transmission time slot allocation with the tracking control constraint, and then performed control optimization based on the specific allocation scheme. In other words, the work also considered the impact of communication on control performance. In \cite{ref15}, the maximum wireless system delay requirements are derived to guarantee the control system stability. Afterwards, the work designed the transmission scheme to meet the control system's delay needs. A very related work \cite{ref16} proposed to use relays to extend the transmission range of platoon messages. The work firstly formulated an integer programming to derive the optimal relay selection with maximizing the link signal quality. Afterwards, a control optimization problem was constructed considering the uncertainty of packet receiving. In words, the routing problem and the control problem were not jointly optimized in \cite{ref16}. Generally speaking, the existing works either study the impact of communication on control performance, either design the communication scheme to meet the control system requirement. 
% These studies demonstrate a consistent methodology: by jointly considering dual requirements and establishing either communication or control constraints based on optimal conditions, they derive optimal communication transmission schemes that satisfy control performance requirements through both single-objective and multi-objective optimization approaches, while deploying optimized control parameters based on communication parameters.
\section{PRELIMINARIES}
\subsection{Notations}
We use uppercase letters to represent matrices and lowercase bold letters to represent vectors\cite{ref17}. We denote the $(i,j)$ entry of a matrix $\boldsymbol{A}$ with $A_{ij}$. We denote a graph $G=\left\{ N,E \right\} $ consisting of $n$ nodes $N=\left\{ n_1,n_2,....,n_n \right\} $ and $l$ edges $E\subseteq N\times N$, where $\left( n_i.n_j \right) \in E$ captures the existence of a link from node $n_i$ to node $n_j$. A graph is connected if an undirected path exists between all pairs of nodes, where each edge $(n_i,n_j)$ is associated with a non-negative weight $w_{ij}\geq 0$. The neighborhood $\mathcal{N}_i$ of node $n_i$ is defined as the set of adjacent nodes ${n_j|(n_i,n_j) \in E }$. The weighted Laplacian matrix $\boldsymbol{L}$ is then defined with off-diagonal entries $L _{ij}=-w_{ij}$ when $\left( n_i.n_j \right) \in E$, zero otherwise, while its diagonal entries satisfy $L_{ij}=\sum_{i\ne j}{w_{ij}}$ to ensure the zero-row-sum property.
\subsection{Control Barrier Function}
\par We consider the following control affine system: 
\begin{equation}
\dot{\boldsymbol{x}}=f\left( \boldsymbol{x} \right) +g\left( \boldsymbol{x} \right) \boldsymbol{u}
\end{equation}
\par \par where state $\boldsymbol{x}\in \mathbb{R} ^n$, input $\boldsymbol{u}\in \mathbb{R} ^m$, and Lipschitz continuous dynamics $f: \mathcal{D}\rightarrow \mathbb{R} ^n$ and $g:\mathcal{D} \rightarrow \mathbb{R} ^{n\times m}$.
\par \textbf{Definition 3.1} \cite{ref18} We define a set $\mathcal{C}$ for the system (1). Let $\mathcal{C}$ be forward invariant, if $\boldsymbol{x_0}\in \mathcal{C}$ implies that $\boldsymbol{x}\in \mathcal{C}$ for all $k\in \mathcal{T}$. Therefore, we call the system (1) safe with the set $\mathcal{C}$.
\par \textbf{Definition 3.2} \cite{ref18} A set $\mathcal{C} \in \mathcal{D}$ is a 0-superlevel set of a continuously differentiable function $h:\mathcal{D} \rightarrow \mathbb{R} $ when
\begin{subequations}
\begin{align}
\mathcal{C} &= \left\{ \boldsymbol{x} \in \mathcal{D} : h\left( \boldsymbol{x} \right) \geqslant 0 \right\}  \\
\partial \mathcal{C} &= \left\{ \boldsymbol{x} \in \mathcal{D} : h\left( \boldsymbol{x} \right) = 0 \right\}  \\
\mathrm{Int}\,\mathcal{C} &= \left\{ \boldsymbol{x} \in \mathcal{D} : h\left( \boldsymbol{x} \right) > 0 \right\} 
\end{align}
\end{subequations}
\par \textbf{Definition 3.3} \cite{ref18} Let $\mathcal{C} =\left\{ \boldsymbol{x}\in \mathbb{R} ^n|h\left( \boldsymbol{x} \right) \geq 0 \right\}$ express as the 0-superlevel set of a continuous function $h:\mathbb{R}^n\rightarrow \mathbb{R} ^n$. The function $h$ is a control barrier function (CBF) for (1) on $\mathcal{C}$ if there exists an $\alpha \in \left[ 0,1 \right] $ for each $\boldsymbol{x}\in \mathbb{R} ^n$, there exists $\boldsymbol{u}\in \mathbb{R} ^n$ such that:
\begin{equation}
h\left( \boldsymbol{x} \right) \geq \alpha h\left( \boldsymbol{x} \right) 
\end{equation}
% \par \textbf{Theorem 1} If $h$ is a DTCBF for (1) associated with \textbf{Definition 2.3}, then any Lipschitz continuous controller $\boldsymbol{u} = k(\boldsymbol{x})$ satisfying
% \begin{equation}
% \nabla h\left( \boldsymbol{x} \right) \left( f\left( \boldsymbol{x} \right) +g(\boldsymbol{x})k(\boldsymbol{x}) \right)+\alpha h\left( \boldsymbol{x} \right) \geq 0
% \end{equation}
% renders $\mathcal{C}$ forward invariant, it ensures $\boldsymbol{x}\left( k_0 \right) \in \mathcal{S} \,\,\Rightarrow \boldsymbol{x}\left( k_t \right) \in \mathcal{S} , \forall k\in I\left( x\left( k_0 \right) \right) $.
% Since the constraint (4) is affine in $\boldsymbol{u}$, it can be incorporated into a quadratic optimization problem: Given a desired control input $k_{des}: \mathcal{D} \rightarrow \mathcal{U} $, we synthesize a safe control input $k: \mathcal{D} \rightarrow \mathcal{U} $ by solving
% \begin{equation}
% \begin{aligned}
% k\left( \boldsymbol{x} \right) = & \underset{\substack{\boldsymbol{x}\in U }}{\arg\min} \,\, \|\boldsymbol{u} - k_{\text{des}}\left( \boldsymbol{x} \right) \|^2 \\
%                     & \text{subj. }\dot{h}\left( \boldsymbol{x},\boldsymbol{u} \right) \geqslant -\alpha \left( h\left( \boldsymbol{x} \right) \right)
% \end{aligned}
% \end{equation}
% in which the nominal control input $\boldsymbol{u}$ is adjusted by the optimizing the problem.
\section{SYSTEM MODEL}
\subsection{Vehicle Model}
We consider a nonholonomic vehicle \cite{ref18} described by 
\begin{equation}
\dot{p}_x=v\,\,\cos \psi , \,\, \dot{p}_y=v\,\,\sin  \psi ,\,\, \dot{\psi}=u_1, \dot{v}=u_2
\end{equation}
\par where $\boldsymbol{p}\triangleq \left[ p_x,p_y \right] ^T$ is the Cartesian coordinates of the vehicle, $\boldsymbol{v}\triangleq \dot{\boldsymbol{p}}$ is the linear velocity, $\psi \in \left( -\pi ,\pi \right] $ is the orientation, and $\boldsymbol{v}\in \mathbb{R} $ is the velocity. The turning rate $r\triangleq \dot{\psi}$ and acceleration $a\triangleq \dot{v}$ are controlled through the inputs $\boldsymbol{u}\triangleq \left[ u_1,u_2 \right] ^T \in U$. The space of admissible inputs is defined by
\begin{equation}
U=\left\{ \boldsymbol{u}\in \mathbb{R} ^2:|r|\leqslant r_{\max},|a|\leqslant a_{\max} \right\} 
\end{equation}
\par where $r_{\max},a_{\max}>0$ represent the maximum rotation rate and acceleration, respectively. We denote that this system can be expressed in the control affine form (1) with $\boldsymbol{x}\triangleq \left[ \boldsymbol{p}^T,\psi ,v \right] ^T \in \mathcal{D}$ and $g=\left[ 0 \,\, \mathbf{I} \right] ^{\mathrm{T}}$ denote the zero and identity matrices, respectively.
\subsection{Velocity Limitation}
\par Rapid velocity fluctuations significantly impact the collision cone. Due to the constraints imposed by multiple driving modes, unpredictable velocity variations introduce substantial safety risks to the system. Therefore, accurately determining vehicle velocity is critical.
\par Each vehicle follows its designated trajectory at velocity $v$. The velocity cones enable to predict and determine collision-free motion constraints. The relative movement between vehicles is defined by velocity vector $\boldsymbol{v}_{j,i}\triangleq \boldsymbol{v}_i-\boldsymbol{v}_j$. If the velocity vector points toward the center of vehicle $j$ within the distance $d_{min,j}$, indicating that vehicle $i$ will lead to a collision with vehicle $j$ if both maintain their velocities with its current trajectory. This critical interaction zone is formally defined as the collision cone , expressed as:
\begin{equation}
\beta_{ij} \triangleq \arcsin \left( \frac{d_{\min,j}}{d_j} \right)
\end{equation}
\par where $d_j\triangleq ||\boldsymbol{r}_j||_2$, $\boldsymbol{r}_j\triangleq \boldsymbol{p}_j-\boldsymbol{p}$ represents the vector starting at the vehicle origin connecting to the center of vehicle $j$. Then the conditions that the velocity cone needs to meet as $\lambda _i<\beta _i$, the angle between $\boldsymbol{v}_{j,i}$ and $\boldsymbol{r}_{j}$ is defined as
\begin{equation}
\lambda _{ij}\triangleq \mathrm{arc}\cos  \left( \frac{\boldsymbol{v}_{j,i}^{T}\boldsymbol{r}_j}{||\boldsymbol{v}_{j.i}||_2d_j} \right) 
\end{equation}
\par \textbf{Lemma 1}\cite{ref18}. Given that vehicle $i$ will not collide with vehicle $j$ at time $\tau \geqslant 0$ for all $\boldsymbol{v}_{j,i}\ne 0$ if
\begin{equation}
\lambda _{ij}\left( k \right) \geqslant \beta _i\left( k \right) , \forall k\geqslant \tau 
\end{equation}
\par The velocity cone serves as a critical safety mechanism. However, there are heterogeneous vehicle types with complex behavioral patterns in complex traffic environments. The limitation of velocity is also decided by the driving operation of the proceeding vehicle $f_i$. To enhance the safety of vehicle, we consider the inter-vehicle distance $\varDelta \boldsymbol{p}_{i{f_i}}\triangleq \boldsymbol{p}_i-\boldsymbol{p}_{f_i}$. Let define the differential inter position\cite{ref20} $\zeta_{i{f_i}}=\left[ \zeta _{p_x},\zeta _{p_y} \right] ^T=\left[ p_{x_i}-p_{x_{f_i}},p_{y_i}-p_{y_{f_i}} \right] ^T$. We also define $D_{i[f_i]}\left( p_{x_i},p_{x_{f_i}} \right) =\sqrt{\left( p_{x_i}-p_{x_{f_i}} \right) ^2+\left( p_{y_i}-p_{y_{f_i}} \right) ^2}$. The operation of vehicle are divided into three typical behaviors: following, braking and changing lanes. Given the inter-vehicle distance $\varDelta {p}_{i{f_i}}$ as the basis for the control barrier function formulation, the corresponding safety function $h_{i{f_i}}$ is defined as
\begin{equation}
\begin{aligned}
&h_{i,f_i}\left( p_i,p_{f_i}\right) = \\& \begin{cases}
	{D_{1}}^2\left( p_i,p_{f_i}\right) -\left( 2W \right) ^2, if\,\,vehicle\,\,follows\,\,others\\
	{D_{2}}^2\left( p_i,p_{f_i}\right) -\left( 2W \right) ^2, if\,\,vehicle\,\,brakes\,\,in\,\,time\\
    {D_{3}}^2\left( p_i,p_{f_i}\right) -\left( 2W \right) ^2, if\,\,vehicle\,\,changes\,\,lanes\\
\end{cases}
\end{aligned}
\end{equation}
\par where $W=\max \left\{ W_1,W_2 \right\}$ represents the safe radius of the safe collision zone, the value of $W_1,W_2$ are determined by the width of lanes. The predicted distance of avoiding collisions is $D_{1}=\sqrt{\left( \varDelta p_{if_i}-l_{i}^{w}\right) ^2}$. The following mode depends solely on the Euclidean distance to the preceding vehicle $f$, $l_{i}^{w}$ is the length of vehicle. $D_{2} =\sqrt{\left(\varDelta p_{if_i} - l_{i}^{w}\right) ^2 + D^1}$, $D^1 = v_i(k) \tau_1 + \frac{v_{i}^2(k)}{2a_{\max}}$,$\tau_1$ represents the response time introduced by the braking action of vehicle $i$. The safe distance of lane-changing is $D_{3} = \sqrt{\left( \varDelta p_{if_i} - l_{i}^{w}\right)^2 + D^2 + D^3}$, $D^2 = v_i(k) \tau_2 + \frac{1}{2} a_{\max} \tau_2^2$ and $D^3 = \frac{\left( v_i(k) + a_{\max} \tau_3 \right)^2 - v_i(k)^2}{2a_{\max}}$. The separation distance $D^1$, $D^2$ and $D^3$ for each segment is computed based on the respective travel distances during different operational phases (including braking deceleration and accelerated lane changes), taking into account the characteristic response times at each stage of vehicle $i$. $\tau_2$ and $\tau_3$ are the response time of two stages in the mode of lane-changing.
\par Subsequently, we analyze the transmission model, taking into account the complex motion dynamics of vehicles within the system, to identify an optimal transmission path that enhances global system performance and ensures reliable velocity information exchange between vehicles.
% \section{PROBLEM FORMULATION}
\subsection{Communication Model}
In this paper, each vehicle can exchange control solutions and data with neighboring vehicles via V2V communication, while simultaneously transmitting service requests to roadside infrastructure through V2I links. The communication path loss between vehicle $i$ and vehicle $j$ \cite{ref21} is intoline-of-sight (LoS) channel, 
\begin{equation}
PL_{i,j}^{\text{LoS}} = 20\log \left( \frac{4\pi f_c}{c} \right) + 20\log \left( \varDelta p_{ij} \right) + \eta_{\text{LoS}}
\end{equation}
\par where $\eta _{LoS}$ is parameters for the LoS channels. $\varDelta p_{ij}$ is the euclidean distance between the nodes, $f_c$ is the carrier frequency and $c$ is the speed of light. Analogously, the V2I path loss between vehicle $i$ and roadside unit (RSU) $I_j$ follows the same propagation model, with the separation distance $\varDelta p_{iI_j}$ defined as the 2D ground projection. The temporal variation in $\varDelta p_{iI_j}$ directly impacts the instantaneous channel conditions.
\par Then, we denote the signal-to-noise ratio (SINR) from two nodes as
\begin{equation}
SINR_{ij}=P^{TX}-PL_{ij}{\text{LoS}}-N_{noise}
\end{equation}
\par where $P^{TX}$ is the power of transmitted signal and $N_{noise}$ is the noise power (dB). Therefore, the transmission rate $R$ of data between two nodes $i$ and $j$ according to the Shannon capacity formula is calculated by
\begin{equation}
R_{ij} = B_{ij} \log_2 \left( 1 + 10^{\frac{SINR_{ij}}{10}} \right)
\end{equation}
\par where $B_{ij}$ denotes the bandwidth of the communication channel of link $l_{ij}$.
% \par The combined effects of channel interference noise and vehicular mobility can lead to intermittent disruptions in data transmission. In this case, node cannot receive all data. We denote the data received by vehicle $i$ from vehicle $j$ as $\varTheta_{ij}$ with the transmission \cite{ref22}. And the set $\varTheta_i=\sum_{j=1}^m{R_{ij}}$ is denoted to be the sum of whole data of the transmission target, we have
% \begin{equation}
% \dot{R}_{ij}\left( t \right) =\begin{cases}
% 	0,   R_i\left( t \right) =\varTheta_i\\
% 	R_{ij},  0\leqslant R_i\left( t \right) <\varTheta_i\\
% \end{cases}
% \end{equation}
\par We establish a minimum SINR threshold $\xi_0 $ for successful vehicular data reception. The transmission is considered successful when the aggregate SINR at receiver $i$ satisfies $\xi_i \geq \xi_0 $, $\xi_i$ denotes the amount of data received by vehicle $i$ in this communication link. Thus, the state model can be obtained as follows \cite{ref18}:
\begin{equation}
x_i\left( k+1 \right) =\boldsymbol{A}x_i\left( k \right) +\boldsymbol{F}u_i\left( k \right) +\delta _{ij}\left( k \right) \cdot \left( x_j\left( k \right) -x_i\left( k \right) \right) 
\end{equation}
\par where $\boldsymbol{A}$ and $\boldsymbol{F}$ are matrices describing the entire system. The probability of $\delta _{ij}$ is calculated by the probability of successfully receiving data $P=Pr\left( \xi_{ij}\geqslant \xi _0 \right)$. If vehicle $i$ successfully receives data of vehicle $j$ at the time slot $k$, let $\delta _{ij}\left( k \right) =1$, otherwise, $\delta _{ij}\left( k \right) =0$. $Pr( \delta _{ij}=1)=Pr\left( \xi_{ij}\geqslant \xi _0 \right),Pr( \delta _{ij}=0)=1-Pr( \delta _{ij}=1)$.
% The system's packet loss rate and recovery rate are defined respectively as $\mathscr{P}$ and $\mathscr{R}$, where $0<\mathscr{P} <1$ and recovery probability $0<\mathscr{R} <1$ are random constants. 

\section{JOINT OPTIMIZATION SCHEME}
\par We propose to design a joint optimization mechanism that optimizes vehicle trajectories while ensuring traffic performance based on control theory, simultaneously incorporate node mobility awareness in routing to enhance transmission stability. This integrated approach will further improve data reliability and consequently refine control performance. The detailed design methodology is presented in the following sections.
\subsection{The Optimal Path of Transmission}
\par The transmission path performance is jointly determined by both the inter-node link quality and the individual node characteristics \cite{ref22}. To comprehensively evaluate the current path's transmission capability, we establish a metric system encompassing node-level and link-level performance indicators.
\subsubsection{Node characteristics of transmission}
\begin{itemize}
\item \textbf{Vehicle status.} The status of vehicle $i$ represents its capability to process the current transfer request. The status is formally defined as $vs_i$, let $vs_i=1$, if $q_i\leqslant q_{\max}$. $vs_i=0$, otherwise. 
% $vs_i=\begin{cases}
% 	1, q_i\leqslant q_{\max}\\
% 	0, otherwise\\
% \end{cases}$, 
$q_i$ indicates the length of the current processing queue of vehicle $i$, and $q_{max}$ indicates the upper limit of the maximum processing queue.
\item \textbf{Number of neighbor vehicles.} Let $\mathcal{N}_i^{path}$ denote the number of neighboring vehicles for vehicle $i$ in this transmission path, while the current transmission performance of vehicle $i$ is evaluated as $\mathcal{N}_i^{path}=\sum_{i=1}^n{vs_i}$.
\item \textbf{Relay reliability.} The relaying reliability $\mathrm{R}_r$ of the $i_{th}$ vehicle quantifies its transmission reliability when performing relaying tasks. Mathematically, it is defined as the ratio between successfully relayed tasks and the total number of tasks received for relaying, expressed as $\mathrm{R}_r=\frac{Nubmer\,\,of\,\,task\,\,relayed}{Total\,\,number\,\,of\,\,task\,\,received\,\,for\,\,relay}$.
% 确定execultion这个数值是否需要
% \item \textbf{Execution reliability.} The execution reliability $\mathrm{R}_e$ of vehicle $i$ is calculated by the same method $\mathrm{R}_e=\frac{The\,\,nubmer\,\,of\,\,task\,\,executed\,\,within\,\,deadline}{The\,\,total\,\,number\,\,of\,\,task\,\,received\,\,for\,\,executed}$.
\end{itemize}
\subsubsection{Link-level characteristics of transmission}
\par  This paper designs a new link metric that fully considers the mobility of vehicles, particularly the impact of their velocity, position, and direction on transmission path vulnerability, disruption, and overhead. The specific link metric is as follows.
\par \textbf{Staying time.} Vehicles can only communicate when they are within each other's transmission range. The residence time is defined as the duration for which vehicle $i$ remains within the communication range of vehicle $j$. Based on their velocity $v_i$ and $v_j$, the duration time of link is denoted as:
\begin{equation}
sd_{j}^{i}=\frac{R_v-\varDelta p_{ij}}{|v_i-v_j|}
\end{equation}
\par where, $R_v$ is communication range of vehicles and $\varDelta p_{ij}$ is distance between two vehicles.
\par \textbf{Direction ratio.} We preferentially select neighbor vehicles with matching movement directions as relay nodes. The direction ratio is given as:
\begin{equation}
d_{\varDelta p_{ij}}=\frac{\varDelta p_{j}^{z}}{\varDelta p_{j}^{s}}
\end{equation}
\par where $\varDelta p_{j}^{z}$ denotes the vector between vehicle $j$ and the destination node 
$z$ based on the difference of spacing distance, while $\varDelta p_{s}^{z}$ denotes the vector between the source node $s$ and the destination node $z$. 
\par \textbf{Velocity variance.} To maintain stable transmission paths with highly dynamic nodes,  we therefore define velocity variance $\sigma _v$ to quantify the degree of variability or dispersion in motion direction and trajectory of vehicles.
\begin{equation}
\sigma _v=\sqrt{\frac{\sum_{j=1,i\in N}^z{\left( v_i-\bar{v} \right)}}{z}}\,\,\,\ \bar{v}=\frac{\sum_{j=1,i\in N}^z{v_i}}{z}
\end{equation}
\par where $\bar{v}$ is the mean of the velocity of vehicles in the path. Vehicle $i$ can only obtain current velocity of vehicle $j$  at time slot $k$ if the transmission succeeds; otherwise, it must rely on the historical velocity of vehicle $j$.
% 针对存在的传输碰撞对齐进行设置指标
\par \textbf{Distance variance.} In wireless network scenarios, as the number of vehicles grows and transmission demands increase within the network system, transmissions from neighboring nodes can cause interference to the current node. We define the probability of successful transmission of V2I or V2V links is expressed as $P_{ij}\left( \theta_{ij} \right)$, packets are transmitted within multiple time slots, $\theta _i\left( k \right)$ represents the amount of data (bits) transmitted by vehicle $j$ to vehicle $i$ in time slot $k$.
\begin{equation}
\begin{aligned}
&prob_{ij}(\theta _i\left( k \right), v_i(k), v_j(k)) =\\& \exp\left(-\frac{2^{\frac{\theta _i\left( k \right)N}{B\tau_{l}}} - 1}{\varpi_{ij}} \cdot L_{ij}^2\left( \dot{v}_{i}\left( k \right) ,v_i(k), ,v_j(k),\dot{v}_{j}\left( k \right)\right)\right)
\end{aligned}
\end{equation}
\par The time-varying relative distance $L_{ij}$ \cite{ref6} between the node $j$ and node $i$, $L_{ij}(v(k),\dot{v}(k))=\sqrt{\left[ p_j(k)-p_i\left( k \right) \right] ^2+L_{0}^{2}}$, $L_0$ is represented the relative vertical distance between vehicles and infrastructures. $L_{ij}$ is related to the control vector and system model. $N$ is the number of vehicles contending the same channel simultaneously. $B$ is the bandwidth of this channel and $\tau_{l}$ is the length of time slot, where $\varpi_{ij}$ represents the normalized power parameters related to SINR. Similarly, the current velocity $v_j(k)$ can only be obtained if the data transmission is successful.
\par If the data is transmitted in multiple time slots (total data volume), the total successful probability of this link is followed as:
\begin{equation}
P_{ij}(\theta _i\left( k \right), v_i(k), v_j(k))= \prod_{k=1}^K prob_{ij}(\theta _i\left( k \right), v_i(k), v_j(k))
\end{equation}
\par \textbf{Path value.} Based on these metrics, we utilize the path metric as an optimization criterion to identify the optimal transmission path. We optimize the transmission path by selecting transmitting nodes based on path value, which evaluates the impact of each candidate node in $\mathcal{N}_i^{path}$. The optimal path is constructed step by step using the highest value at each node. The value is calculated as follows:
\begin{equation}
\varDelta _{path}=\frac{sd^{i}_{j} \times w_i \times d_{\varDelta p_{ij}}}{\sigma _v}\times P_{ij}(\theta _i\left( k \right), v_i(k), v_j(k))
\end{equation}
\par where $w$ is the weight of node $i$ is set by node characteristic. The weight of node $i$ is calculated as:
\begin{equation}
w_i = 
\begin{cases}
    \kappa_1 \cdot vs_i + \kappa_2 \cdot \mathcal{N}_i^{path} \\ 
        \quad + \kappa_3 \cdot \mathrm{R}_{r}^{i}, & 
        \text{the node is able to transmit}, \\
    1, & \text{otherwise}. \\
\end{cases}
\end{equation}
\par where $\kappa_1,\kappa_2$ and $\kappa_3$ respectively represent the weight values corresponding to thees features of each node.
\subsection{The Control Scheme}
\par Channel interference, link congestion, timeout and so on will cause communication problems, resulting in packet loss, in this paper we use the Markov packet loss model framework to accurately capture the stochastic nature of the packet loss process \cite{ref21}. However, due to the problem of data packet loss existing in the communication transmission process, the vehicle need to ensure the safety of vehicle formation control and the stability of vehicle formation regardless of whether it can successfully receive the data packet at present. Based on this, we construct and design the following the control objective. We denote $T$ to be the predictive horizon length. Let $K=\left\{ 0,1,2,....,T \right\} $. We define the cost function $J$ as determined by the control input $\boldsymbol{u}$, the deviation from the ideal trajectory $\dot{x}$, and the state $\boldsymbol{x}$ deviation among neighboring vehicles \cite{ref23}. The cost function of vehicle $i$ is defined as
\begin{equation}
\begin{aligned}
&J_i=||u_{i}\left( k\right) ||_{R_i}+||\dot{x}_{i}\left( k\right) -x_i\left( k\right) ||_{F_i}\\&+\sum_{j\in N}{||\dot{x}_{i}\left( k\right) -x_j\left( k\right) -\varDelta {p}_{ij}||_{G_i}}
\end{aligned}
\end{equation}
\par where $R$ and $F$ are the positive definite matrix, $G$ is the symmetric matrix. $x_i\left( k\right)$ represents the state of the control input predicted by vehicle $i$ at time slot $k$. The predicted trajectory is denoted by $\dot{x_{i}}\left( k \right)$ with the predicted control input $\dot{u_{i}\left( k\right)}$ and the initial state $x_i(k)$, where $x_{i}\left( 0\right) =x_i\left( k \right) $.
% \par Then an optimization objective of vehicle $i$ at time $K$ is designed as follows.
% \begin{equation}
% \begin{aligned}
% \min_{u_i^p (k|t), k \in \mathbb{K}_0} & J_i \left( x_i^p, u_i^p, x_i, x_j \in N; \mathbb{K}_0 | k \right) \\
% &= \sum_{k=0}^{T_p-1} l_i \left( x_i^p, u_i^p, x_i, x_j \in N; k | K \right)
% \end{aligned}
% \end{equation}
\par However, vehicle $i$ cannot maintain seamless communication with all neighboring vehicles at all times. During the time interval $[K-1,K]$, vehicle $i$ continuously receives data until the transmission is fully completed. At this stage, vehicle $i$ cannot acquire the most recent data and must rely solely on historical data from the time interval $[0,K-1]$. To further mitigate the impact of unreliable data transmission on control performance, we define the self-deviation constraint as \cite{ref23}:
\begin{equation}
\begin{aligned}
\gamma_i \sum_{k=1}^{T_p-1} \| \dot{x_i (k)} - x_i(k) \|_{G_i} \\\leq \sum_{k=1}^{T-1} \| \dot{x_i(k-1)} - x_i(k-1) \|_{G_i}
\end{aligned}
\end{equation}
\par Then the optimization problem \cite{ref23} is designed as:
\begin{subequations}
\begin{equation}
\min_{u_i(k),k\in\mathbb{K}_0} J_i = \sum_{0}^{K-1} J_i(k)+ \delta_{ij}(k)\sum_{K-1}^{K} J_i(k) 
\end{equation}
\begin{equation}
\text{s.t.} \quad \left\{
\begin{aligned}
&x_i(k+1) = \boldsymbol{A}x_i(k) + \boldsymbol{F}u_i(k) \\ 
&\qquad + \delta_{ij}(k) \cdot (x_j(k) - x_i(k)), \\
&\gamma_i \sum_{k=1}^{T-1} \|\dot{x}_i(k) - x_i(k)\| \\
&\qquad \leq \sum_{k=1}^{T-1} \|\dot{x}_i(k-1) - x_i(k-1)\|, \\
&a_{\min} \leq \bar{v}\leq a_{\max}, \,\, v_{\min}\leq v\leq v_{\max} \\
&\psi_{\min} \leq \psi \leq \psi_{\max} ,\,\, r_{\min} \leq r \leq r_{\max}\\
&h_{i,j}\left( p_i,p_j \right) \geqslant -\alpha \left( h_{i,j} \right), \forall i,j\in N \\
&\lambda _{ij}\left( k \right) \geqslant \beta _{ij}\left( k \right) , \forall k\geqslant \tau, \forall i,j \in N
\end{aligned}
\right. 
\end{equation}
\end{subequations}
\par where $\gamma_i$ is a scale coefficient. These constraints are initial state, control input constraints, the vehicle dynamics constraints, CBF constraints and self-deviation constraint to ensure system stability under random packet loss.
\subsection{The Transmission Optimization Scheme}
% \par Based on the above analysis, packet transmission over any link $l\in \mathcal{L} $ occupies exactly one time slot using one of $C$ available orthogonal channels in communication network, with the interference property that concurrent transmissions on distinct channels remain non-interfering while same-channel transmissions may collide. Within the network topology $G$, multiple packets of $J$ distinct types (denoted $J=\left\{ 1,2,...,J \right\} $) are transmitted concurrently through V2V and V2I links, with each packet type $m\in J$ characterized by its source node $s_m\in N$, destination node $z_m\in N$, deadline offset $o_m\in \mathbb{N} \cup \left\{ 0 \right\} $. Packets of type $m$ needing to arrive at time slot $t$ must be delivered before slot $t+o_m$, otherwise being discarded. Let denote $o_{max}$ as the maximum deadline offset and $\gamma _min$ as the minimum arrival rate, the effective network operation horizon extends to $\bar{T}:=T+o_{\max}$ to accommodate all packets. Successful delivery requires establishing an interference-free transmission path comprising a sequence of properly scheduled link-channel pairs that guides packets from $s_m$ to $z_m$ within their respective deadlines $t+o_m$.
\par The data packet transmission optimization can be formulated as an integer programming problem over the time horizon $T_p$. The optimization \cite{ref24} involves two key decision variables the packet scheduling $Z_{\ell}^{mk}$ and routing decisions $y_{l}^{gk}$, where $y_{l}^{gk} \in {0,1}$ indicates whether the arriving packet is scheduled using relative path $l$ at time $k$, $l$ represents the communication link of node $i$ and node $j$. The channel activation variables which represent the network scheduling decisions.
\begin{subequations}
\begin{align}
&\max \sum_{k=0}^K \sum_{g=1}^{\varTheta_{k}} \sum_{\ell \in L} y_{l}^{gk}\varDelta _{path} \\
&\text{s.t.} \quad \left\{
\begin{aligned}
& \sum_{\ell \in L} y_{l}^{gk}\leq 1, \\
& \sum_{\tau=k-k_{max}}^{\tau \wedge K} \sum_{g=1}^{\varTheta_{k} ^\tau} \sum_{l: l_{k-\tau} = \ell} y_{l}^{g\tau} \leq \sum_{m=1}^M \sum_{l: \ell \in l} Z_{\ell}^{mk}, \\
& \sum_{\ell \in L} Z_{\ell}^{mk} \leq 1, \quad \forall m \in [M], k \in [K], \ell \in [L]\\
& Z_{\ell}^{mk} \in \{0,1\}, \quad \forall m \in [M], k \in [K], \ell \in [L] \\
& y_{l}^{gk} \in \{0,1\}, \quad \forall k \in [K], l \in [L].
\end{aligned}
\right. 
\end{align}
\end{subequations}
\par where $Z_l^{mk}$ indicates that at each time and channel, one independent set is selected for the network schedule. $\varTheta_i$ is denoted to be the sum of data of the transmission target from node $i$. $m$ and $g$ represent the number of channel and the number of packets respectively. The term $(k-k_{max})$ denotes the remaining time slots from the arrival slot $k$ until the deadline for a packet. This constraint guarantees that for any time slot and arriving packet, the total scheduled transmissions of packet on link $l$ during the interval $[k, k+k_j]$ cannot exceed the number of channels allocated to link $l$.
\subsection{The Joint Optimization}
\par To achieve optimal data transmission in vehicular networks, the platoon must account for both the complexity of vehicle dynamics and the challenges posed by unreliable communication channels. More precise vehicle trajectory prediction enables the system to select optimal transmission paths, while transmission optimization significantly reduces packet loss probability. Consequently, we formulate a dual-objective optimization DynaRoute framework that simultaneously: (1) maximizes transmission performance and (2) minimizes formation oscillation. Building upon this foundation, we propose a joint optimization model for vehicular communication networks. This comprehensive model integrates key aspects of vehicle mobility, formation control theory, and communication reliability, formally expressed as follows:
\begin{equation}
\begin{aligned}
\max_{u,v,p} & \quad Y(u,v,p) \\
           \min_{u} & \quad J(u) \\
\text{s.t.} & \quad 
\begin{cases}
\begin{aligned}
&x_i(k+1) = \boldsymbol{A}x_i(k) + \boldsymbol{F}u_i(k) \\
&\qquad + \delta_{ij}(k) \cdot (x_j(k) - x_i(k)), \\
&h_{i,j}\left( p_i,p_j \right) \geqslant -\alpha \left( h_{i,j} \right), \forall i,j \in N \\
&\lambda _{ij}\left( k \right) \geqslant \beta _{ij}\left( k \right) , \forall k\geqslant \tau, \forall i,j \in N;\\
&u_i(k) \geq 0, \quad i \in N, k \in K\\
&\delta_{ij}(k) \in \{0,1\}.
\end{aligned}
\end{cases}
\end{aligned}
\end{equation}
% \section{The algorithm}
\par We convert the minimization problem $\underset{u}{\min}: J\left( u\right)$ to the minimization problem $\underset{u}{\max}: -J\left( u\right)$. Next, we calculate the maximum optimization objective of the new optimization $Y\left( u,v,p \right)-J(u)$. The optimization problem is solved by Nondominated Sorting Genetic Algorithm(NSGA) \cite{ref25}. 
\subsection{Iterative Optimization Algorithm}
\par This paper proposes a collaborative transmission optimization mechanism based on the improved NSGA-II algorithm and DMPC controller alogrithm \cite{ref23}. While ensuring reliable platoon control performance, optimization is conducted in the communication link selection with desire position $p^*$, velocity $v^*$ and acceleration $a^*$. The solution of (23) guides vehicles for optimal control  performance. We transform the optimization problems into maximization problem. The process is as follows.

\begin{algorithm}
\caption{Distributed Model Predictive Control for Vehicle Platooning}
\begin{algorithmic}
    \State \textbf{Input:} 
    \begin{itemize}
        \item Initialization state $x_0$
        \item Maximum calculation times $CalMax$
        \item Vehicle constraints of $u,v,p$
    \end{itemize}
    \State \textbf{Output:} Pareto-optimal solutions $u^*$
    \State \textbf{Initialization of the leader vehicle} $k = 0$: \textbf{for} $i \in N$ \textbf{do}
    \State a) Prearrange $x_i(0)$ and $u_i(0)$:
    \[
    x_i(0) = 
    \begin{cases} 
    x_i(0), & k = 0, \\
    A x_i(k-1) + F u_i(k-1), & k \in K. 
    \end{cases}
    \]
    \[
    u_i(0) \in U_i, \quad k \in K.
    \]

    \State \textbf{Online Phase at following vehicle} $k \geq 0$: \textbf{for} $i \in N$ \textbf{do}
    \State a) Convey $x_i(k)$ to vehicles $j \in N$;
    \State b) Receive $x_j(k)$ from vehicles $j \in N$;
    \State c) Resolve the probability using Markov states, then attain $u_i^*(k)$ and $x_i^*(k)$;
    \State d) Resolve the probability of $\delta_{ij}$, determine whether neighbor states are successfully acquired;
    \State e) Take $u_i^*(0)$ for the control of vehicle $i$;
    \State f) Prearrange $x_i^*(k+1)$ and $u_i^*(k+1)$:
    \[
    x_i^*(k+1) =
    \begin{cases} 
    x_i^*(k), \text{if vehicle failed}, \\
    A x_i(k) + F u_i(k) + \delta_{ij}(k) \cdot (x_j(k) - x_i(k)),  \\ 
    \text{if vehicle succeeded}.
    \end{cases}
    \]
    \[
    u_i^*(k+1) =
    \begin{cases} 
    u_i^*(k),  \text{if vehicle failed}, \\
    \sum_{j \in N} \delta_{ij}(k) \left( x_j(k) - x_i(k) - \Delta p_{j,i} \right), \\
    \text{if vehicle succeeded}.
    \end{cases}
    \]

    \State \textbf{Iteration at } $t \geq 0$: \textbf{for} $i \in N$,
    \State \quad a) Set initial solution and candidates:
    \[
    \mathrm{solution}_i(0) \gets \emptyset, \quad \mathrm{candidates}_i(0) \gets \text{getInitialCandidates}()
    \]
    \State \quad b) Select candidate with maximum value:
    \[
    u^*_i(t) \gets \underset{c \in \mathrm{candidates}_i(t)}{\arg\max} \, J(c)
    \]

    \State \quad c) Update solution if feasible:
    \begin{align*}
    \mathrm{solution}_i(t+1) \gets 
    \begin{cases}
    \mathrm{solution}_i(t) \cup \{u^*_i(t)\} \\ \text{if } \mathrm{isFeasible}(u^*_i(t)), \\
    \mathrm{solution}_i(t),  \text{otherwise}.
    \end{cases}
    \end{align*}

    \State \quad d) Prune candidates:
    \begin{align*}
    \mathrm{candidates}_i(t+1) \gets 
    \begin{cases}
    \mathrm{candidates}_i(t) \setminus \{u^*_i(t)\} \\ \text{if } \mathrm{isFeasible}(u^*_i(t)), \\
    \mathrm{candidates}_i(t),  \text{otherwise}.
    \end{cases}
    \end{align*}

    \State \quad e) Check termination:
    \[
    \text{If } \mathrm{isTerminal}(\mathrm{solution}_i(t+1)) \text{, return } \mathrm{solution}_i(t+1).
    \]

\end{algorithmic}
\end{algorithm}

% 增加顺利接收和不顺利接收时，状态参数的变化过程
\par According to the aforementioned algorithm, the optimal solution is obtained as $u^* = \mathop{\arg\min}\limits_{u \in \mathcal{U}} J(u)$. The solution of (25) for a set of path values under the current node's motion based on $p^*$, $v^*$ and $a^*$, determining the selection for each hop. Therefore, at the timestep $k$, for the current node $i$, the next-hop node $j$ and transmission channel $g$ are selected based on the value of optimization in (25) derived from a set of hop-by-hop determined $y^{gk}_l$. Through multiple iterations, we optimize and determine the desired solution and select the maximum value as the optimal solution by calculating the value of optimization in (25). 
\par To address the two joint optimization objectives, the NSGA-II algorithm is employed to solve the joint multi-objective optimization problem. The detailed algorithm is as follows: 
\par Step 1: We first define an empty set $\mathcal{Y}_0=\left\{ y_{l}^{gk}=0|k\in K,l\in L,g\in \left\{ 0,1,2,...,\varTheta \right\} \right\}$ to store the resulting candidate solutions $y^{gk}_l$, where $l \in [L]$ at initial iteration $0$. We acquire the current network topology in VANET, we first construct a link set derived from the neighborhood relationships between the source and destination nodes. By traversing the entire dynamic topology, we generate a collection of transmission decision schemes. And then we store the set of solutions in collection $\mathcal{Y}_0=\left\{ y_{l}^{gk}(0)=0|k\in K,l\in L,g\in \left\{ 0,1,2,...,\varTheta \right\} \right\}$, thereby obtaining the initialized set $\mathcal{Y}_t=\left\{ y_{l}^{gk}(t)|k\in K,l\in L,g\in \left\{ 0,1,2,...,\varTheta \right\} \right\}$ at iteration $t$. 
\par Step 2: We employ a set of predefined reference points to ensure diversity in the obtained solutions. Based on three indicators as $u,v,p$, each indicator is partitioned into $10$ segments according to reference location coordinates, adaptively generating $55$ reference points\cite{ref27}. The algorithm for selecting reference points is as follows:
\par Let $C$ be a two-dimensional variable defined over the domain ${0,\frac{1}{10},\frac{2}{10},...,\frac{11}{10}}$, where $c_{ij}$ represents the evenly distributed parts of the three indexes of $u,v,p$. For each $c_{ij}\in C$, $c_{ij}=c_{ij}-\frac{j-1}{10}$. We rank the individuals within $\mathcal{Y}_t=\left\{ y_{l}^{gk}(t)|k\in K,l\in L,g\in \left\{ 0,1,2,...,\varTheta \right\} \right\}$ based on reference points to effectively enhance solution diversity.
\par Step 3: We perform non-dominated sorting on $\mathcal{Y}_t=\left\{ y_{l}^{gk}(t)|k\in K,l\in L,g\in \left\{ 0,1,2,...,\varTheta \right\} \right\}$. Based on the calculations of two optimization functions in (25), the corresponding values are obtained and assigned. Then the set $\mathcal{Y}_t=\left\{ y_{l}^{gk}(t)|k\in K,l\in L,g\in \left\{ 0,1,2,...,\varTheta \right\} \right\}$ undergoes non-dominated sorting and crowding distance calculation to select parents, followed by crossover and mutation operations to generate suitable offspring. Then we construct the new set $\mathcal{Y}^{'}=\left\{ y_{l}^{gk}(t)|k\in K,l\in L,g\in \left\{ 0,1,2,...,\varTheta \right\} \right\}$. The basic steps are as follows:
\par We select individuals $y_{l_1}^{gk}(t)$ and $y_{l_2}^{gk}(t)$ from $\mathcal{Y}_t=\left\{ y_{l}^{gk}(t)|k\in K,l\in L,g\in \left\{ 0,1,2,...,\varTheta \right\} \right\}$ after sorted with lower non-dominated ranks and higher dynamic crowding distances through tournament selection, then generate offspring $y_{l_1}^{gk}(t+1)$ and $y_{l_2}^{gk}(t+1)$ of the iteration $t$ to proceed to compute and update.
% these solutions:
% $$
% y_{l_1}^{gk}(t+1)=0.5\left[ \left( 1+\varphi _i \right) y_{l_1}^{gk}(t)+\left( 1-\varphi _i \right) y_{l_2}^{gk}(t) \right] 
% $$
% $$
% y_{l_2}^{gk}(t+2)=0.5\left[ \left( 1-\varphi _i \right) y_{l_1}^{gk}(t)+\left( 1+\varphi _i \right) y_{l_2}^{gk}(t) \right] $$.
\par Step 4: According to the above mutation calculations, we obtain the new values of offspring are obtained and stored in the set $\mathcal{Y}^{'}$ and we construct the set $\mathcal{Y}_{t+1}=\mathcal{Y}_t \cup \mathcal{Y}^{'}_t$, where $\mathcal{Y}_{t+1}=\left\{ y_{l}^{gk}(t+1)|k\in K,l\in L,g\in \left\{ 0,1,2,...,\varTheta \right\} \right\}$. The values retained from the previous two steps and the newly calculated values are then merged to form the set $\mathcal{Y}_{t+1}$ for the next iteration $t+1$.
\par Step 5: For the iteration $t+1$, we first judge whether the algorithm has met the termination condition. If $t+1$ exceeds the maximum iteration count $GenMax$, terminate the algorithm and output the non-dominated solutions in $\mathcal{Y}_t=\left\{ y_{l}^{gk}(t)|k\in K ,l\in L,g\in \left\{ 0,1,2,...,\varTheta \right\} \right\}$ as the Pareto optimal set, the optimal transmission path is determined based on the corresponding optimal values following the calculation of the value of optimization in (25) and then we select the maximum value. The transmission path scheme corresponding to the maximum value is then identified as the optimal solution, we then get the best value of the set $y^{gk}_l$; otherwise, continue executing the NSGA-II algorithm from the above steps until we choose the optimal transmission.
% 简易复杂度分析和简易收敛性分析
The algorithm is shown as follows.
% 伪代码中的参数表达需要再次对应
\begin{algorithm}
\caption{NSGA-II for Multi-Objective Optimization}
\begin{algorithmic} % 关键修改：[0]表示禁用行号
\State \textbf{Input:} 
    \begin{itemize}
        \item Population size $N$
        \item Maximum generations $GenMax$
        \item The maximum solution of $J(u^*)$
    \end{itemize}
\State \textbf{Output:} Pareto-optimal solutions $y^*$

% \State Initialize population $P_0$ with $N$ individuals parents $y^{gk}_l$
% \State Evaluate objectives by (25) for each individual
% \State Perform non-dominated sort on $Y_0=\left\{ y_{l}^{gk}=0|k\in K,l\in L,g\in \left\{ 0,1,2,...,\varTheta \right\} \right\}$ to assign ranks
% \State Compute crowding distance for diversity preservation to reduced to one-fourth of $Y_0=\left\{ y_{l}^{gk}=0|k\in K,l\in L,g\in \left\{ 0,1,2,...,\varTheta \right\} \right\}$

\For{$t \gets 1$ to $GenMax$}
    % \State \textbf{Reproduction:}
    % \State Select parents using binary tournament selection
    % \State Apply simulated binary crossover (SBX) with probability $p_c$
    % \State Apply polynomial mutation with probability $p_m$
    % \State Source node and destination node
    
    % \State \textbf{Evaluation:}
    % \For{each offspring $y(u,v,p)$ in $Y^{'}=\left\{ y_{l}^{gk}(t)|k\in K,l\in L,g\in \left\{ 0,1,2,...,\varTheta \right\} \right\}$}
    %     \State Compute $Y(u^*,v^*,p^*) \gets max Y(u,v,p)$ in (25)\Comment{Maximization objective}
    %     \State Compute $J(u^*) \gets max -J(u)$ in (25)\Comment{Minimization transformed}
    %     \State \textbf{Check constraints:}
    %     \If{any of $u$, $v$, $p$ violated}
    %         \State Assign penalty fitness values
    %     \EndIf
    % \EndFor
    
    % \State Combine parent and offspring populations $Y_t=Y_t \cup Y^{'}_t$
    % \State Perform non-dominated sort on combined population
    % \State Select new population $Y_{t+1}$ based on ranks and crowding distance
    \State \textbf{Initialization at } $t = 0$:
\State \quad a) Generate initial population:
\[
\mathcal{Y}_t(i) \gets \text{randomIndividual}(), \quad \mathcal{Y}^{'}_t \gets \emptyset
\]
\State \quad b) Evaluate the first objective and constraints:
\[
\max_{u,v,p} \quad -J(u) , \quad \forall u \in U
\]
\State \quad c) Evaluate the second objective and constraints:
\[
\max_{u,v,p} \quad Y(y^{gk}_l) , \quad \forall k \in K, l \in L, g\in \left\{ 0,1,2,...,\varTheta \right\}
\]
\State \textbf{Generational loop at } $t \geq 0$:
\State \quad a) Combine populations:
\[
\mathcal{Y}_t \gets \mathcal{Y}_t \cup \mathcal{Y}^{'}_t
\]
\State \quad b) Non-dominated sorting:
\[
(y^{gk}_{l_1}, y^{gk}_{l_2}, \dots) \gets \text{sortNonDominated}(\mathcal{Y}_t)
\]
\State \quad c) Crowding distance calculation with the first objective:
\begin{itemize}
    \item $\text{Distance}(x) \gets \sum_{k=1}^m \frac{Y(y^{gk}_{l_i}) - Y(y^{gk}_{l_{i+1}})}{Y^{\max} - Y^{\min}}$
    \item $\forall k, l, g$
\end{itemize}
\State \quad d) Crowding distance calculation with joint objective:
\begin{itemize}
    \item $\text{DISTANCE}(x) \gets \sum_{k=1}^m \frac{-J(u_i) + J(u_{i+1})}{(-J^{\max}) - (-J^{\min})}$
    \item $\forall u \in U$
\end{itemize}
\State \quad e) Crowding distance calculation with the second objective:
\[
\text{distance}(x) \gets \text{Distance}(x)-\text{DISTANCE}(x), 
\]
\State \quad f) Population selection:
\[
\mathcal{Y}_{t+1} \gets \bigcup_{i=1}^{l} y^{gk}_{l_i} \quad \text{until } |\mathcal{Y}_{t+1}| + |\mathcal{Y}_{l+1}^{'}| > N
\]
\State \quad g) Generate offspring:
\[
\mathcal{Y}_{t+1}^{'} \gets \text{tournamentSelect}(\mathcal{Y}_{t+1}) \oplus \text{adaptive crossover/mutation}
\]
\State \quad h) Termination:
\[
\text{If } t+1 = GernMax \text{, return } \text{ParetoFront}(\mathcal{Y}_t).
\]
\EndFor
\end{algorithmic}
\end{algorithm}

\section{PERFORMANCE EVALUATION}
\par This section presents a comprehensive evaluation of the proposed dual-objective joint optimization model through systematic simulation studies. First, we provide a detailed description of the simulation environment and parameter configurations. Subsequently, we present an in-depth analysis of the optimization model's performance metrics. Finally, to rigorously validate the proposed approach, we conduct comparative experiments assessing both V2I communication reliability and platoon control stability. All numerical simulations were executed using python. The hardware configuration features an Xeon W-2135 processor (3.7 GHz base frequency, 64-bit architecture) with 32GB RAM, ensuring robust computational capability for the simulation scenarios.
\subsection{Simulation Setup}
\par The simulation environment comprises two parallel straight lanes populated with multiple Connected and Autonomous Vehicle (CAVs). In this setup, we deploy two distinct platoons, each consisting of four vehicles (one leader and three followers) operating on separate lanes, totaling eight CAVs in the simulation. We establish the vehicle mobility parameters are configured following the methodology in \cite{ref2}, while the V2I communication parameters are implemented according to \cite{ref3}. For precise initialization, Table 1 comprehensively documents the complete parameter set governing fleet mobility dynamics and physical layer communication characteristics.

\begin{table}[!ht]
\centering
\caption{Parameters Settings in Model}
\begin{tabular}{|l|l|l|l|}
\hline
\textbf{Parameter} & \textbf{Value} & \textbf{Parameter} & \textbf{Value} \\ \hline
% $N$               & 4,8             & $N_{Access}$               & 150           \\ \hline
% $K$               & 30 s          & $\tau$            & 0.5 s         \\ \hline
$\epsilon$        & $3 \times 10^{-5}$ & $d_i$        & 10 m          \\ \hline
$a_{\min}$        & $-2.5 \, \text{m/s}^2$ & $a_{\max}$ & $+2.5 \, \text{m/s}^2$ \\ \hline
$\xi^{v}_{\min}$    & $-6 \, \text{m/s}$ & $\xi^{v}_{\max}$ & $+6 \, \text{m/s}$ \\ \hline
$\xi^{p}_{\min}$    & $-3 \, \text{m}$ & $\xi^{p}_{\max}$ & $+3 \, \text{m}$ \\ \hline
$\varTheta_i$ & $5 \times 10^9 \, \text{bits}$ & $B$ & 10 Mbit \\ \hline
$\varpi$          & $-96 \, \text{dBm}$ & $K$               & 30 s          \\ \hline
$L_0$             & $50 \, \text{m}$ & $L_{V2I,0}$ & $200 \, \text{m}$ \\ \hline
\end{tabular}
\end{table}
\par The scenario evaluates the resilience of connected vehicle systems. The simulation employs a $0.1s$ sampling interval for platoon control. We predefine the leading vehicle's acceleration and deceleration profiles, with detailed velocity variation patterns. The accelerations and decelerations of the lead vehicle is shown as:
\begin{equation}
a_0\left( k \right) = \begin{cases}
    0.5 \, \text{m/s}^2, \quad k \in \left[ 3.5, 5.5 \right] \, \text{s} \\
    1 \, \text{m/s}^2, \quad k \in \left[ 6.0, 7.5 \right] \, \text{s} \\
    0.5 \, \text{m/s}^2, \quad k \in \left[ 8.0, 10.0 \right] \, \text{s} \\
    -0.5 \, \text{m/s}^2, \quad k \in \left[ 14.5, 16.5 \right] \, \text{s} \\
    -1 \, \text{m/s}^2, \quad k \in \left[ 17.0, 18.5 \right] \, \text{s} \\
    -1 \, \text{m/s}^2, \quad k \in \left[ 19.0, 21.0 \right] \, \text{s} \\
    0 \, \text{m/s}^2, \quad \text{others}
\end{cases}  
\end{equation}

\subsection{Experimental Results}
% \par Experiment 1 studies the transmission performance under the joint optimization scheme, investigates the time dynamics of the successful transmission probability under different time slot durations, and analyzes the evolution of this key performance index under different configurations. The cumulative data transmission volumes in the simulation were systematically evaluated to characterize the throughput efficiency of the system.
\par Experiment 1 examined the transmission performance of two mechanisms in VANET. The first method is the joint optimization method designed in this paper. The second method is the Artificial Intelligence-aware Emergency Routing Protocol. This method constructs a Belief-Desire-Intention (BDI) model to understand the environment, integrating data such as traffic density, congestion levels, collision signals, and hazard detection to optimize route selection for the system. We have respectively implemented the intelligent transportation system routing protocols based on decentralized Model Predictive Control (DMPC) \cite{ref21} and artificial intelligence (AI) \cite{ref26} of Emergency routing protocol (ERP) in the vehicle queue as reference comparisons. 
\begin{figure}[htbp] % h:此处 t:顶部 b:底部 p:单独页
  \centering
  \includegraphics[width=0.8\linewidth]{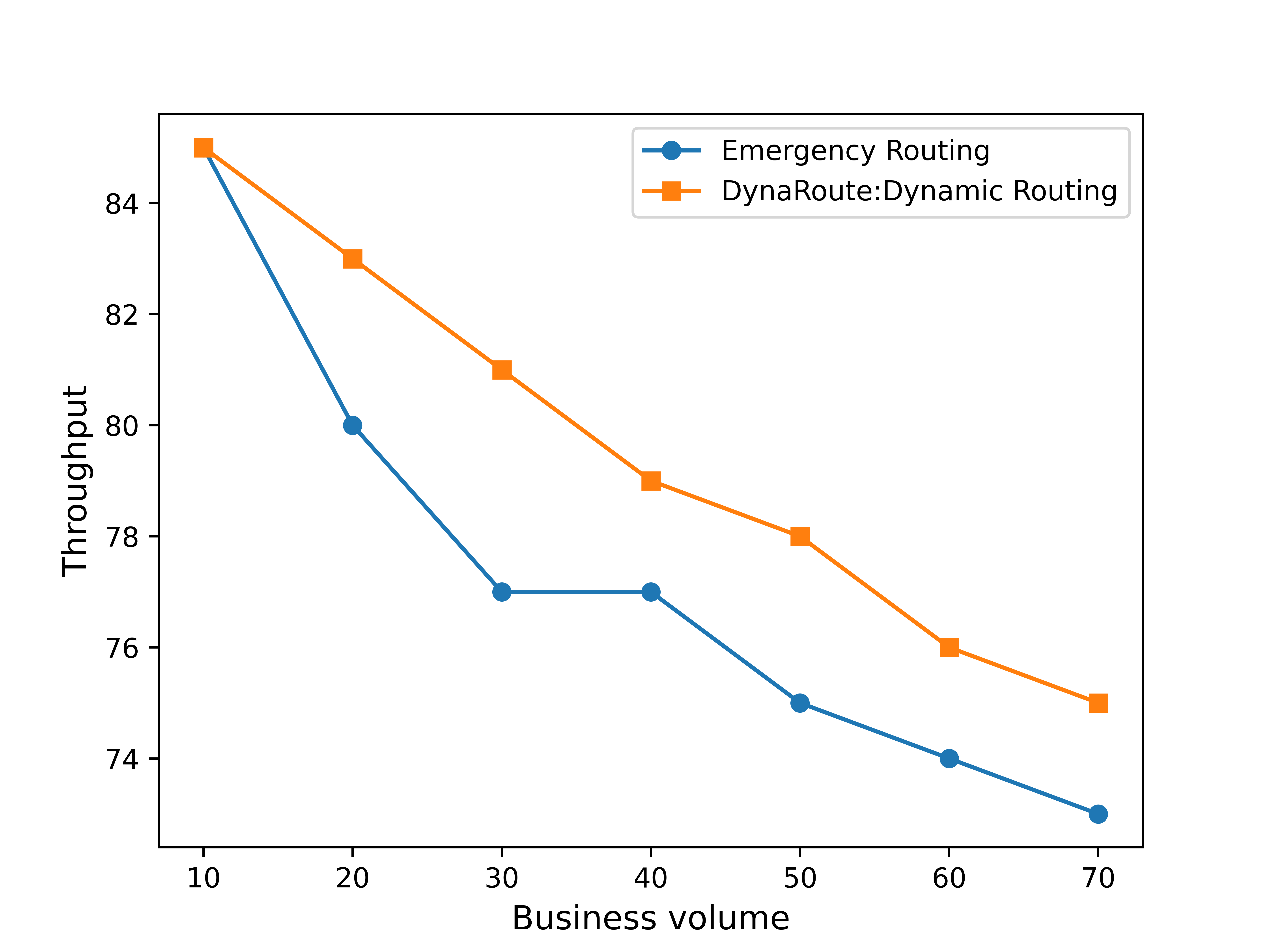} % 宽度设为栏宽的80%
  \caption{Throughput under different network load levels }
\end{figure}
\begin{figure}[htbp] % h:此处 t:顶部 b:底部 p:单独页
  \centering
  \includegraphics[width=0.8\linewidth]{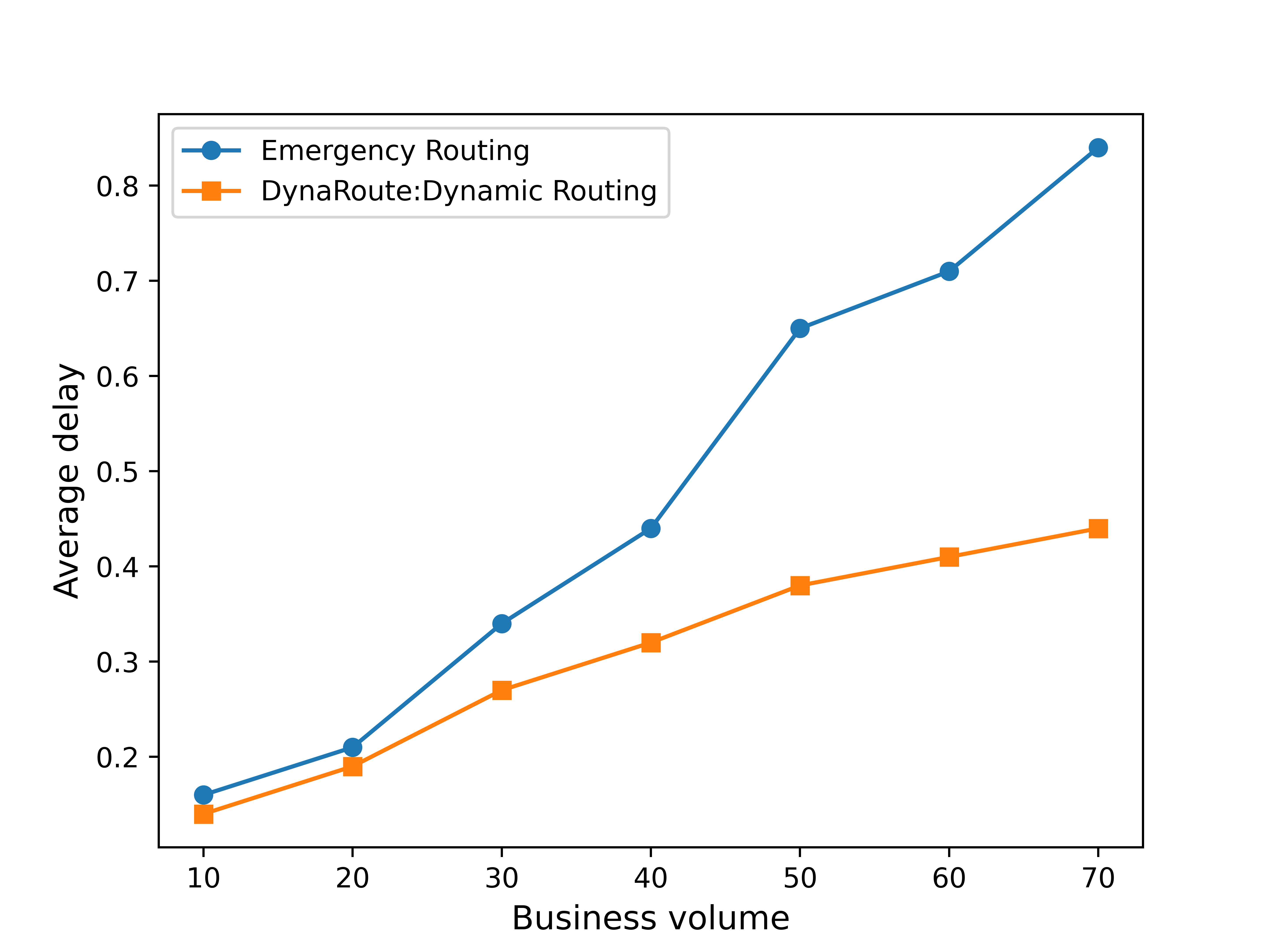} % 宽度设为栏宽的80%
  \caption{End to end delay under different network load levels}
\end{figure}
\begin{figure}[htbp] % h:此处 t:顶部 b:底部 p:单独页
  \centering
  \includegraphics[width=0.8\linewidth]{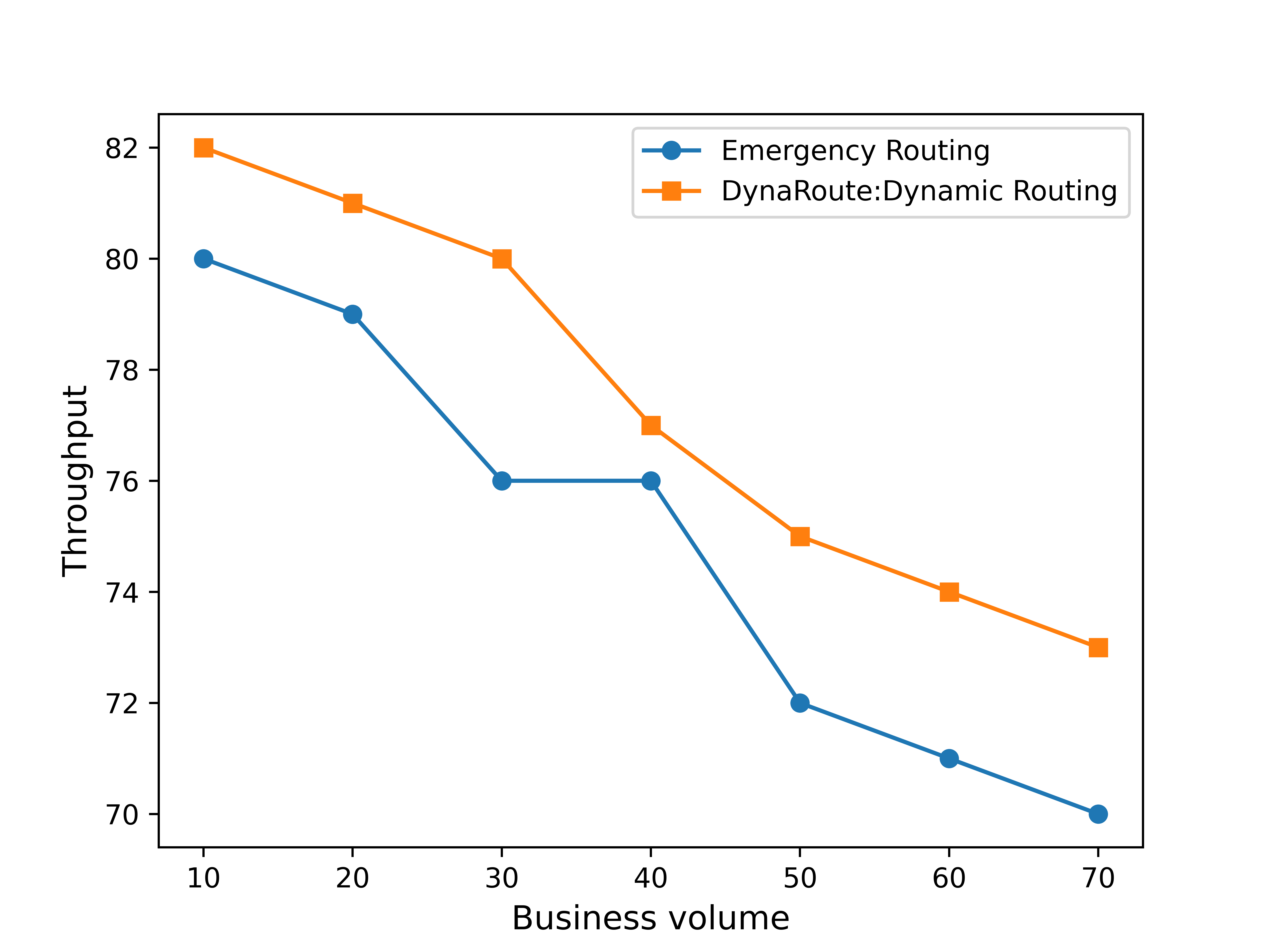} % 宽度设为栏宽的80%
  \caption{Throughput under different packet interval levels}
\end{figure}
\begin{figure}[htbp] % h:此处 t:顶部 b:底部 p:单独页
  \centering
  \includegraphics[width=0.8\linewidth]{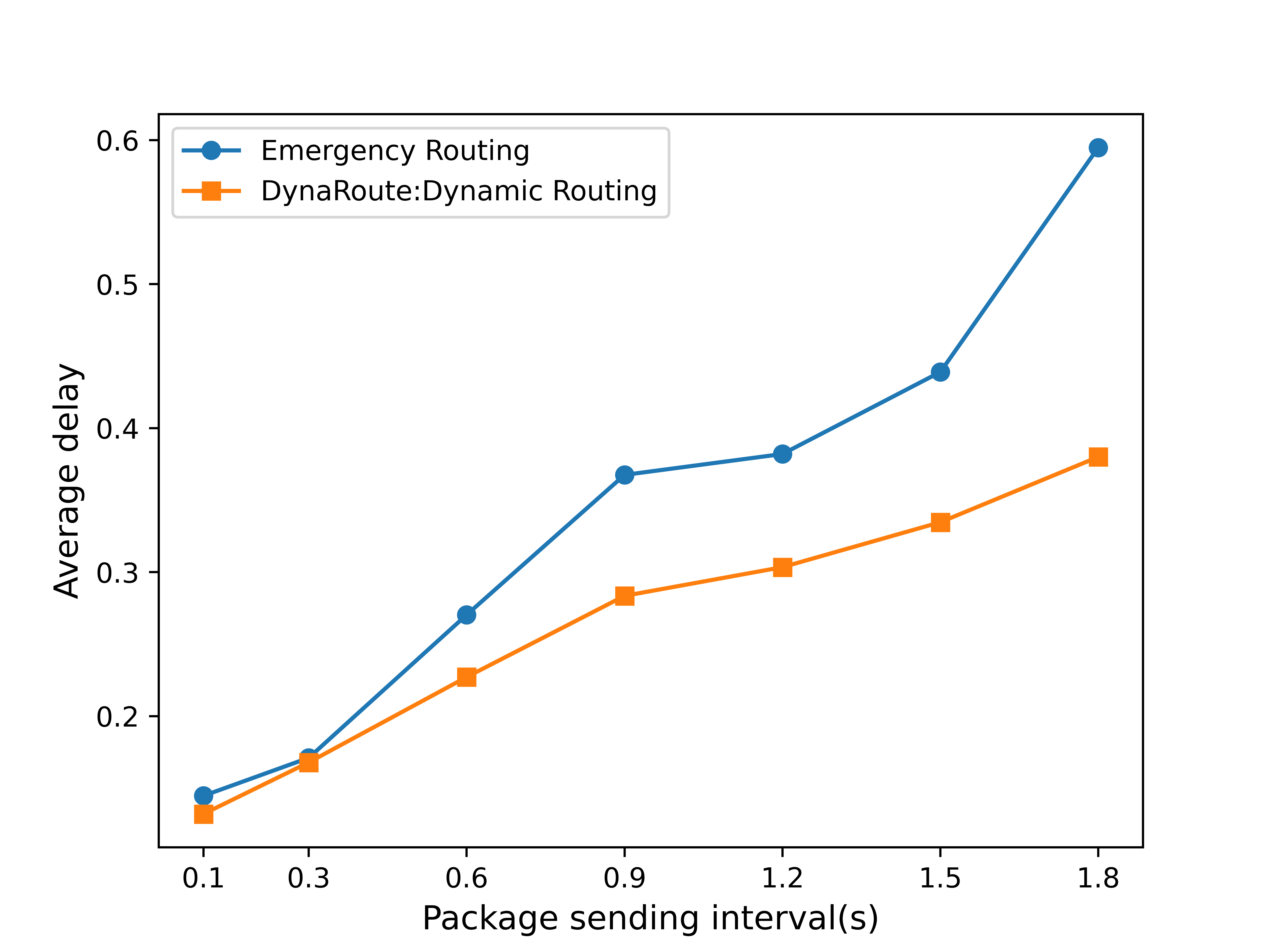} % 宽度设为栏宽的80%
  \caption{End to end delay under different packet interval levels}
\end{figure}

\par As shown in Fig. 1-4, the horizontal axes represent network load and packet transmission interval, while the vertical axes include throughput and end-to-end delay. These figures illustrate the transmission performance curves of the vehicular network under different network loads and different packet interval levels. In Fig. 1, when the load is light, two mechanisms maintain satisfactory transmission performance. As the network load increases with the number of packets, the rising transmission delay raises end to end delay, leading to a significant performance degradation in ERP. As depicted in Fig. 3, as the number of data packets transmitted in the network increases, node load becomes heavier and transmission performance gradually deteriorates. The ERP fails to adequately account for dynamic link metrics, resulting in significant performance degradation. The joint optimization mechanism dynamically adjust link metrics, identifying optimal links that meet multiple criteria under varying network loads. Fig. 2 and Fig. 4 compare the packet transmission delays of the two mechanisms under different packet interval, ERP does not fully account for node mobility, leading to increased link fragility in complex vehicle movement scenarios, which slightly degrades its performance. As the load grows, this contention increases packet loss rates and substantially raises end-to-end delay. In contrast, the proposed mechanism consistently prioritize the most stable paths for vehicles, maintaining superior transmission performance despite load fluctuations caused by traffic variations. The results demonstrate that the proposed mechanism achieves stable transmission performance under varying network load conditions, exhibiting strong robustness against both node mobility and dynamic network loads.
\par Experiment 2 examined the impact of the joint mechanism on the dynamics of platoon formation. Under the condition of leading vehicle dynamics changes, the study analyzed the guidance of the joint optimization mechanism in this paper, taking into account the influence of communication interference between vehicles on the reception of control information such as velocity of neighbour vehicles. We investigated the time-varying patterns of the trajectories, vehicle distances, vehicle velocities and control inputs (acceleration) of eight vehicles within two platoons.
% 参考之前车排控制的仿真文字描述

\begin{figure*}[t] % 使用figure*跨双栏
    \centering
    \begin{minipage}[t]{0.32\textwidth}
        \includegraphics[width=\linewidth]{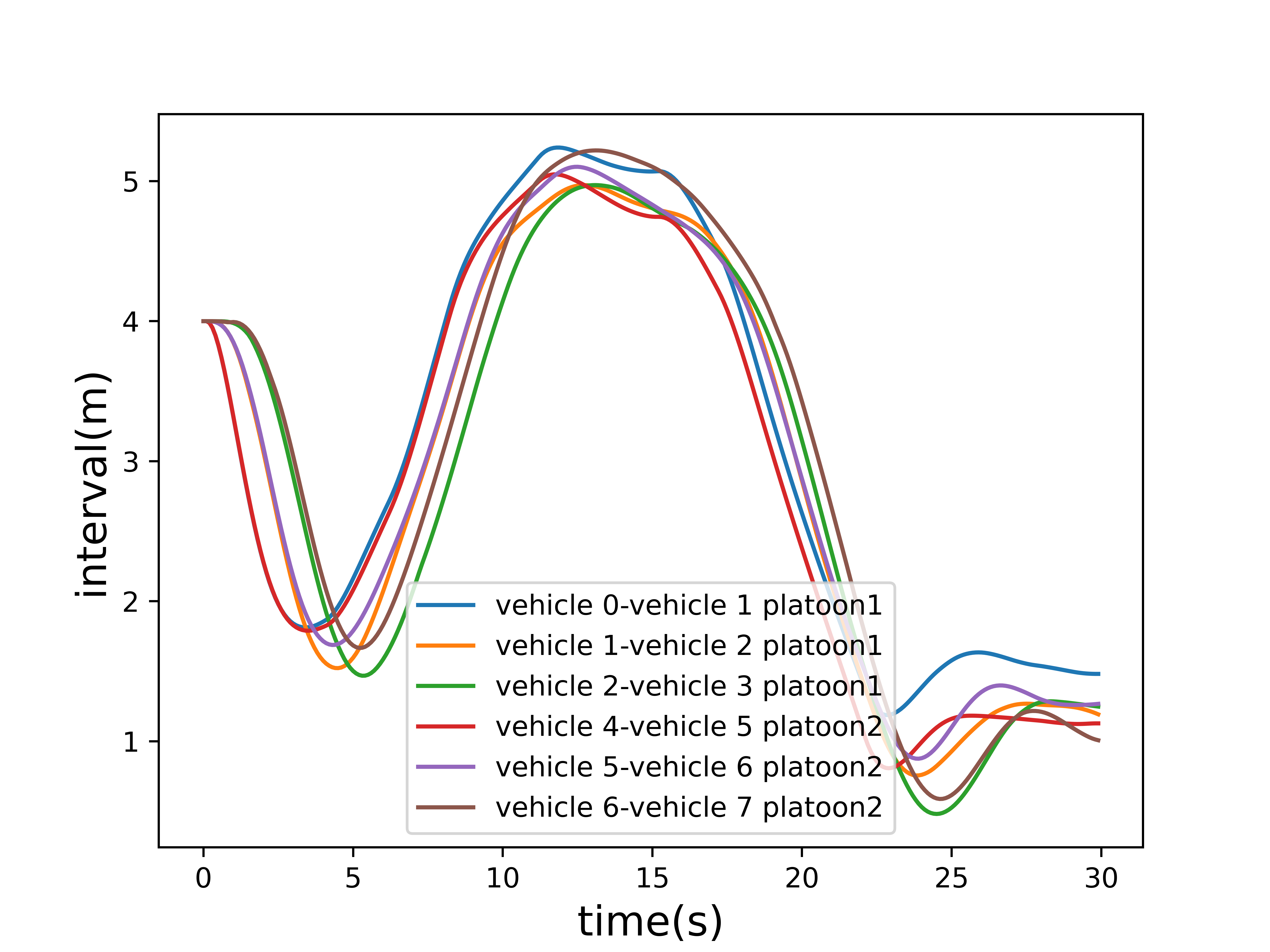}
        \caption{interval distance in case1}
    \end{minipage}
    \hfill
    \begin{minipage}[t]{0.32\textwidth}
        \includegraphics[width=\linewidth]{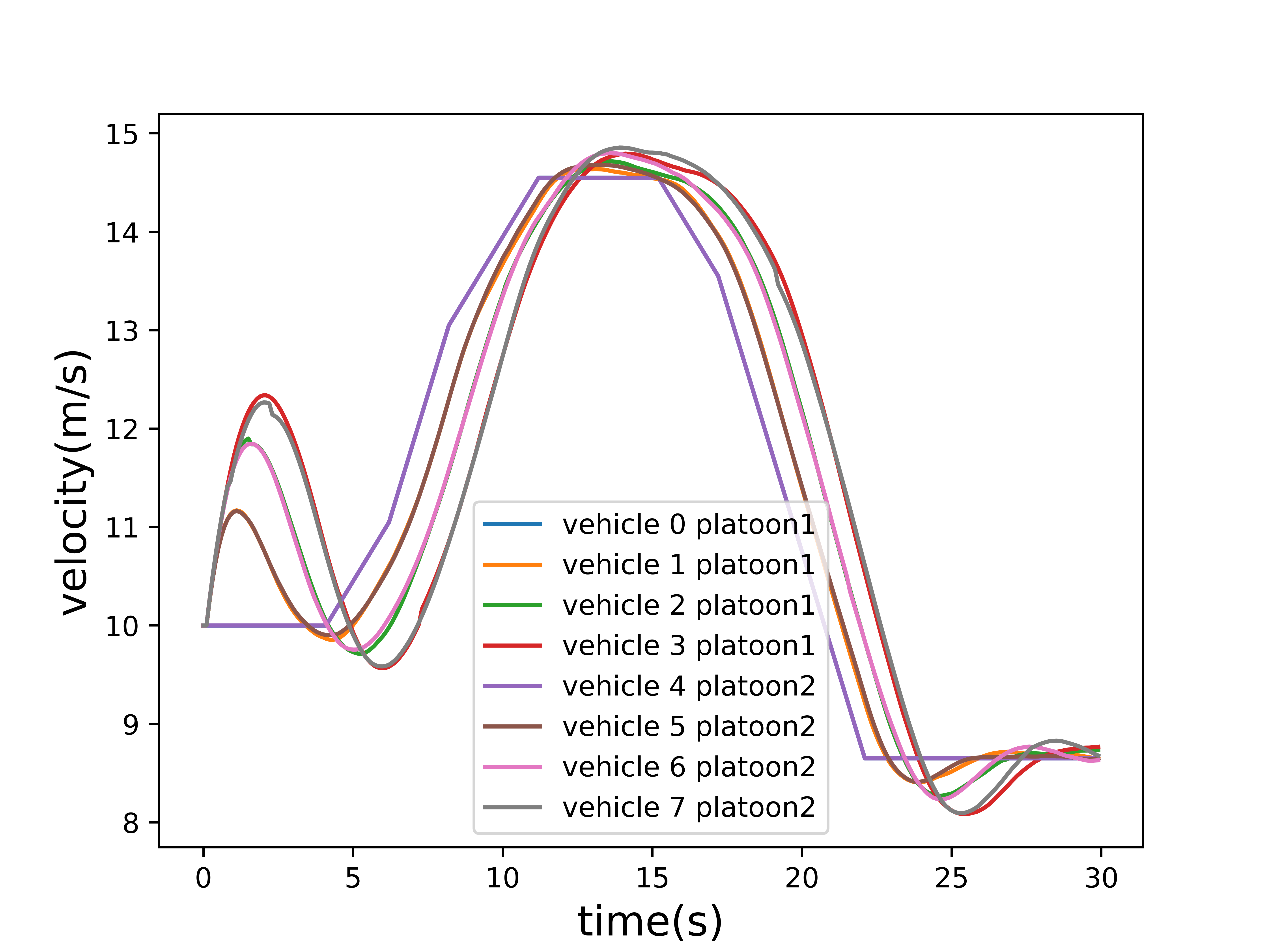}
        \caption{velocity in case1}
    \end{minipage}
    \hfill
    \begin{minipage}[t]{0.32\textwidth}
        \includegraphics[width=\linewidth]{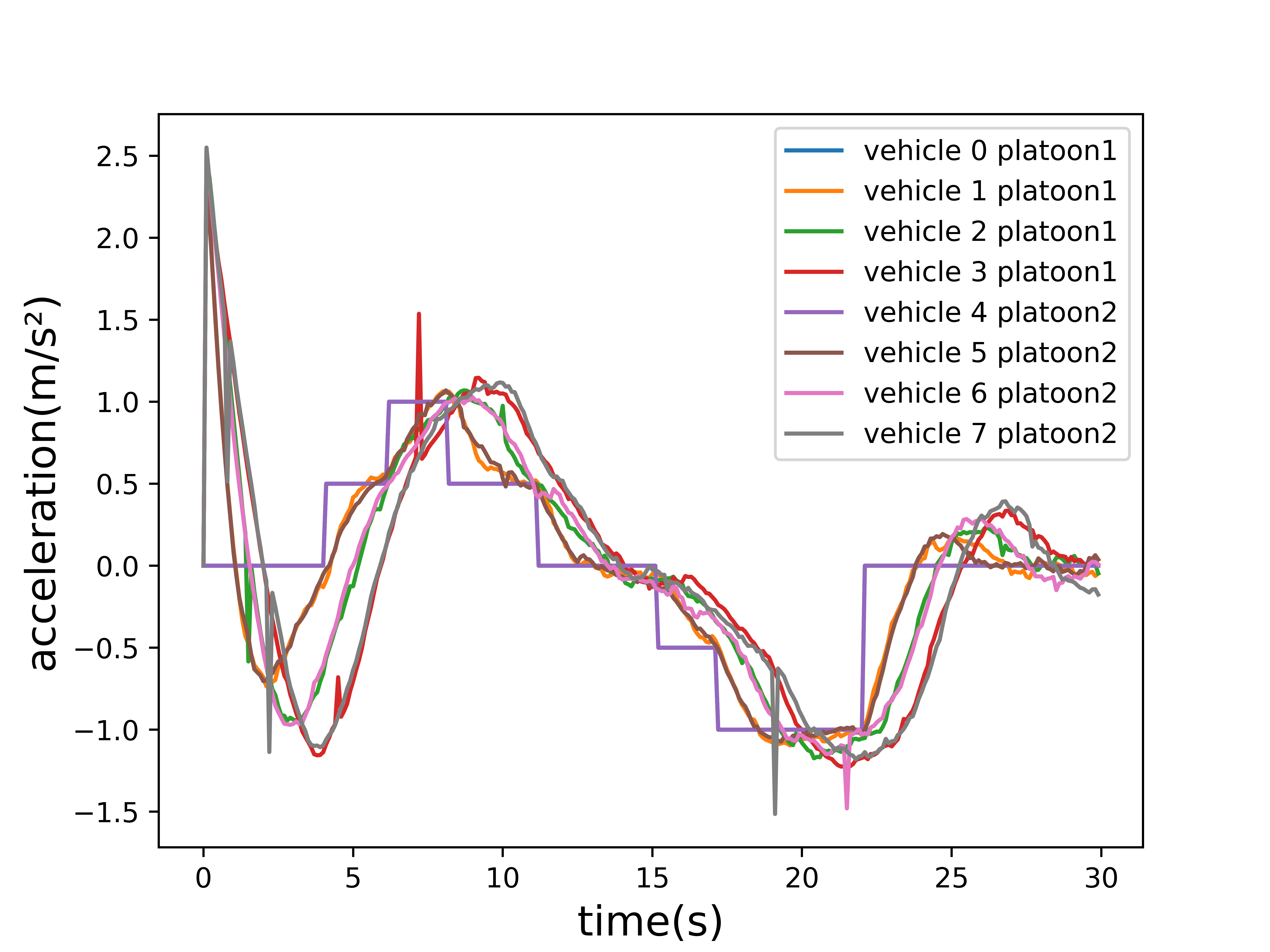}
        \caption{acceleration in case1}
    \end{minipage}
    % 将主标题放在所有子图之后
    % \caption{Figure_11}
\end{figure*}

\begin{figure*}[t] % 使用figure*跨双栏
    \centering
    \begin{minipage}[t]{0.32\textwidth}
        \includegraphics[width=\linewidth]{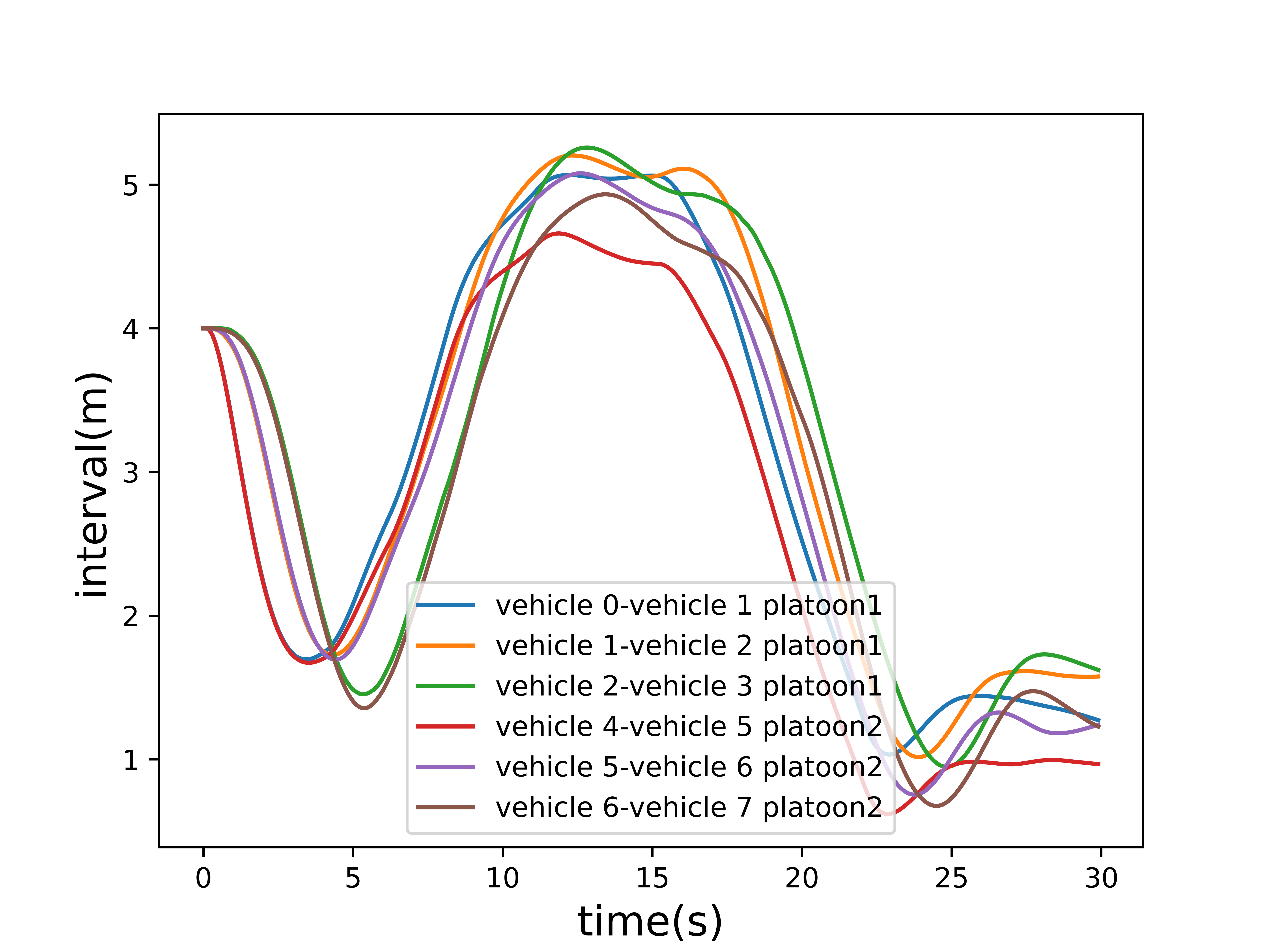}
        \caption{interval distance in case2}
    \end{minipage}
    \hfill
    \begin{minipage}[t]{0.32\textwidth}
        \includegraphics[width=\linewidth]{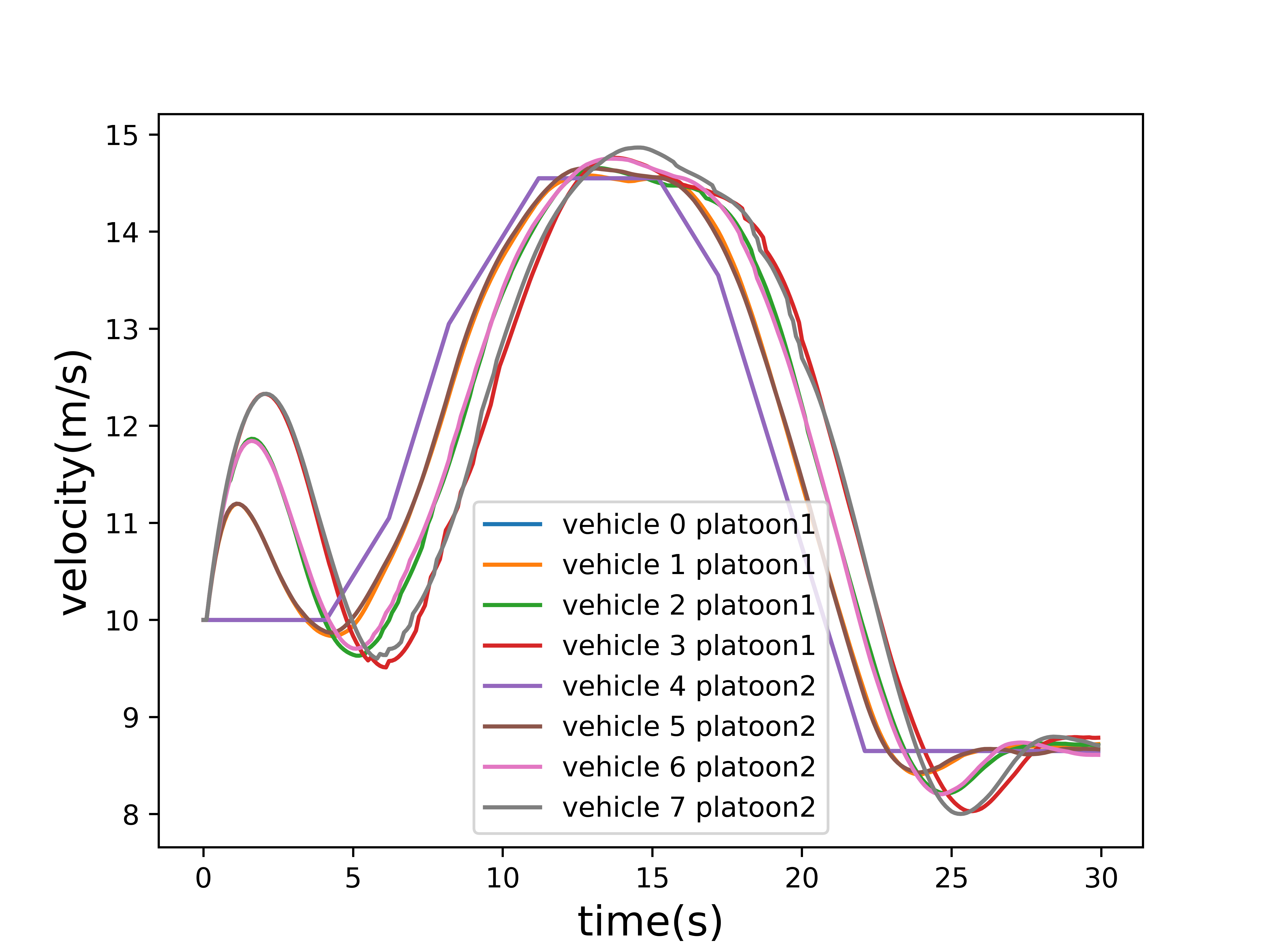}
        \caption{velocity in case2}
    \end{minipage}
    \hfill
    \begin{minipage}[t]{0.32\textwidth}
        \includegraphics[width=\linewidth]{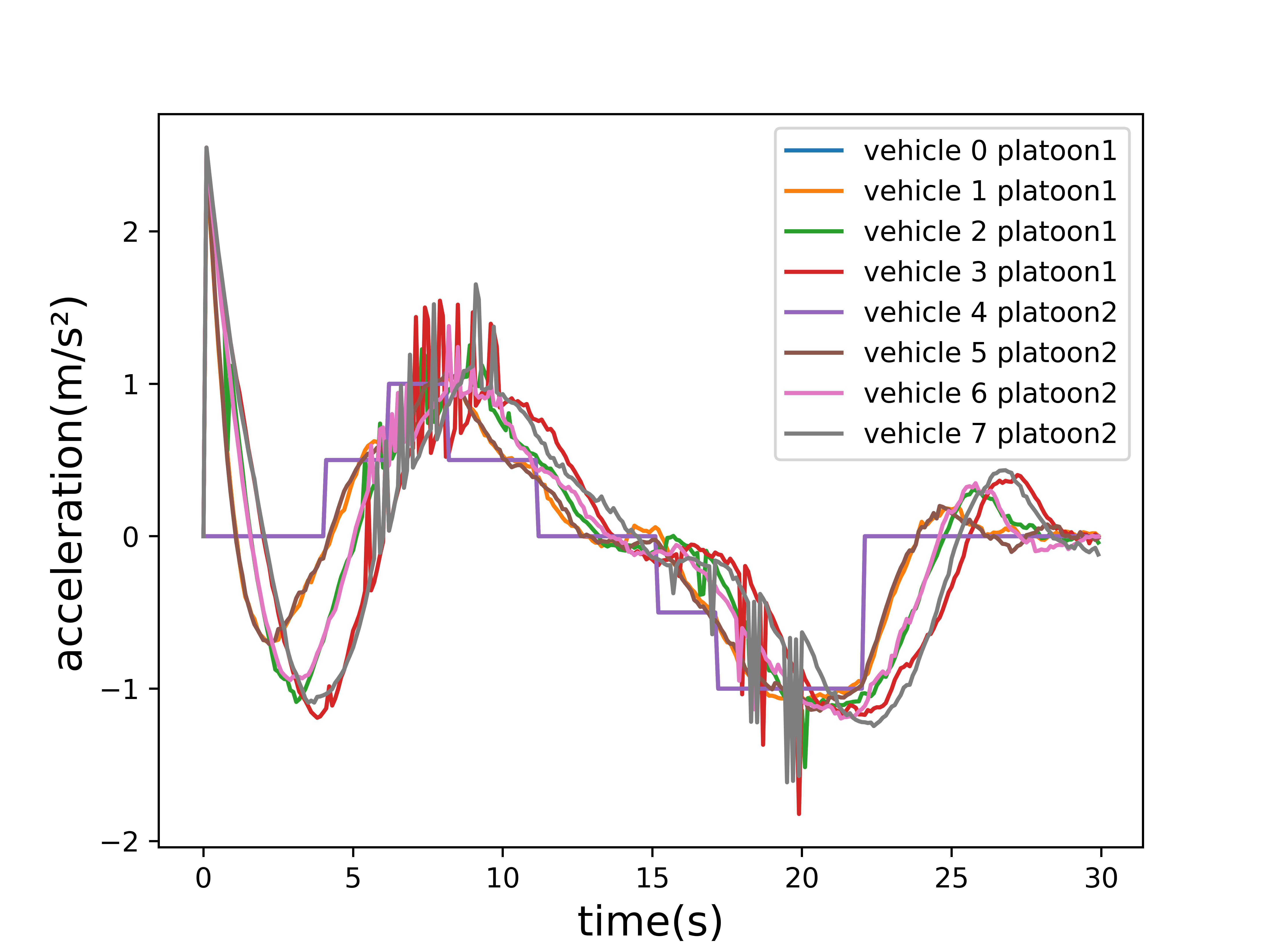}
        \caption{acceleration in case2}
    \end{minipage}
    % 主标题应该放在所有子图之后
    % \caption{Figure_12}
\end{figure*}

\par The two sets of figures illustrate that in both packet-loss scenarios for vehicular networks, vehicles require control mechanisms to compensate for errors and interference caused by transmission failures. Fig. 5-7 demonstrate the control performance under network packet loss conditions (case1), while Fig. 9-11 show the control performance in overloaded vehicular network systems with dense vehicle presence and severe packet transmission obstacles (case2). Fig. 5-11 present the real-time position, velocity, and acceleration of eight vehicles in two platoons under the joint optimization scheme proposed in this paper. As seen in Fig. 5 and Fig. 9, the inter-vehicle distance consistently remains greater than zero, effectively eliminating collision risks. Regardless of irregular velocity variations in the leading vehicle, all following vehicles maintain a safe following distance quickly through an appropriate response mechanism. The results in Fig. 6 and Fig. 10 reveal that in both platoons, vehicles 0-7 exhibit rapid velocity responses to changes in preceding vehicles. According to the definition of string stability in \cite{ref29}, both platoons maintain stability under these dynamic conditions. Since vehicle acceleration is determined by control inputs, it proves most sensitive to variations. The acceleration curves in Fig. 7 show significant correlation with velocity values of adjacent vehicles, with all disturbance fluctuations strictly bounded within $\pm 1 m/s^2$. The joint control mechanism effectively mitigates interference from packet loss and other transmission errors while consistently maintaining efficient platoon control performance. In Fig. 11, despite severe packet loss and fluctuating transmission performance, acceleration variations intense yet the fluctuations of disturbance are also bounded with  $\pm2 m/s^2$. The control input always preserve safe driving. These empirical parameter variations demonstrate that the proposed mechanism can reliably guide vehicles to operate safely and efficiently in vehicular networks, even under potential information loss or transmission delays.

\section{CONCLUSION}
\par We introduce DynaRoute, a new transmission optimization mechanism that leverages control mechanisms and precise node mobility prediction in Internet of Vehicles scenarios to simultaneously ensure safe vehicle guidance and maintain stable, reliable communication links for high-velocity mobility. DynaRoute enhances network throughput while minimizing link disruptions and handoffs caused by node movement. The mechanism employs computing to jointly solve control and transmission optimization problems, determining optimal solutions for multi-objective challenges and identifying ideal transmission paths for source-destination pairs. Our simulation-based evaluation demonstrates that the performance of DynaRoute, whether measured by throughput, end to end delay, or driving safety metrics remains robust against common wireless network impairments including interference, channel contention, and packet loss.

% \include{reference.tex}

% \vspace{12pt}
% \color{red}
% IEEE conference templates contain guidance text for composing and formatting conference papers. Please ensure that all template text is removed from your conference paper prior to submission to the conference. Failure to remove the template text from your paper may result in your paper not being published.

\end{document}